\documentclass[12pt]{iopart}
\usepackage{graphicx}
\usepackage{iopams}

\expandafter\let\csname equation*\endcsname\relax
\expandafter\let\csname endequation*\endcsname\relax
\usepackage{amsmath}
\usepackage{mathtools}
\usepackage [autostyle, english = american]{csquotes}
\usepackage{xcolor}
\begin{document}

\title[]{Study of two-electron temperature plasma sheath using nonextensive electron distribution in presence of an external magnetic field}

\author{G. Sharma$^{1}$, R. Paul$^{1}$, K. Deka$^{1}$, R. Moulick$^{1}$, S. Adhikari$^{2}$, S. S. Kausik$^{1,*}$, and B. K. Saikia$^{1}$}

\address{$^1$Centre of Plasma Physics, Institute for Plasma Research, Nazirakhat, Sonapur-782402, Kamrup(M), Assam, India}
\address{$^2$Department of Physics, University of Oslo, PO Box 1048 Blindern, NO-0316 Oslo, Norway}
\address{$^*$E-mail: kausikss@rediffmail.com}
\vspace{10pt}

\begin{abstract}
In this study, the physics of sheath formation in a collisional two-electron
temperature plasma in the presence of an oblique external magnetic field has been investigated. At first, a comparative study among the fluid electron
model, Boltzmann electron model and the nonextensive electron model has been carried out and a suitable range of
nonextensive parameter $q$ has been predicted. In the latter part, a collisional two-electron
temperature plasma is considered. Both the hot and cold electron densities are
described using the non-extensive distribution whereas cold ions are described by the
fluid equations. The properties of the sheath are investigated in different collisional regimes
by varying the non-extensive parameter ($q$) and the hot to cold electron densities and
temperatures. The magnetic field inclination angle is varied in the limit $1^0 \leq \alpha \leq 5^0$. It is observed that electron distribution significantly deviates from Boltzmann distribution for nearly parallel magnetic field. Moreover, collision enhanced flux deposition for highly magnetised case is a significant finding of the study. The results obtained in this study can enhance the understanding of plasma matter interaction processes where multiple electron groups with near parallel magnetic field are found.
\end{abstract}

%
%
%
%
%

\section{Introduction}
The non-neutral region present in the plasma-wall interface is known as the plasma sheath, which determines the plasma surface interaction processes. Due to the non-neutral nature of the region, there exists a strong electric field, which controls the ion flux to the wall. Therefore, sheath study continues to draw reasonable attention of the plasma processing industry as well as the fusion research community for many decades \cite{KU}. The behavior of this space charge dominated region depends on the ambient plasma properties. In a collisionless unmagnetized plasma, the ion dynamics are controlled by the local electric field and the well-known Bohm criterion is fulfilled at the sheath entrance \cite{Bohm}. But in collisional plasma, the Bohm criterion does not have to necessarily be fulfilled and ions may enter into the sheath at a subsonic speed \cite{Val}. Further, in the presence of an oblique magnetic field, the structure of the sheath utterly changes, and accordingly, the Bohm criterion is also generalized \cite{Chodu}. The problem becomes even more complicated and interesting in multi-component plasmas. In plasmas with two species of positive ions, the  electrostatic sheath potential is considerably affected by the heavier ion species in the presence of ion-neutral collision \cite{Hatami1}.  In collisional low-temperature plasmas, the electrons have a much higher temperature in comparison to the ions. More often, non-equilibrium stationary states of electron distribution are observed in laboratory plasmas \cite{Stan, Sheridan,Ike, My_2}. In such cases, the total electron density can be divided into two components, \textit{viz.}, hot and cold electrons. Such plasmas are termed as two-electron temperature plasmas. Recent studies have shown that the presence of an energetic electron group has a predominant effect on the plasma dynamics \cite{Shukla, Gyr, Ou, My}. In these studies, it is assumed that Boltzmann-Gibbs (BG) statistics is valid for such systems and hence the electrons are described using the Boltzmann distribution. However, the electron velocity distribution may readily deviate from the Boltzmann distribution in astrophysical as well as laboratory plasmas. In the magnetospheres of the earth and the Saturn, the electron velocity distribution is found to be well described by the kappa distribution \cite{Vasy, Koen}. Moreover,  many experimental studies have reported the occurrence of non-Maxwellian electron distribution in various laboratory conditions \cite{Jaw, Kakati, Kalita, Chen, Liu}. Hence, Maxwellian assumption does not always hold good at some of the situations.

It is well known that BG statistics is valid for macroscopic equilibrium states and suitable for describing short-range particle interactions. However, long-range interactions are quite common in plasmas. Therefore, the Boltzmann distribution can not adequately describe all the plasma dynamics. In the recent past, a new generalized statistical description has been put forward by Tsallis, known as Tsallis distribution or non-extensive distribution \cite{Tsa}. It is found that this distribution is capable of describing the systems that deviate from the regular Maxwell-Boltzmann distribution. Since, in the presence of energetic electrons and external magnetic fields, the electrons no longer remain Boltzmann distributed \cite{Fran, Tsha}, therefore, this new generalized distribution might be more suitable in this regard. In fact, many potential researchers have carried out dynamical studies considering non-extensive electron distribution \cite{Hatami2, Safa, QH, Bas}. Borgohain \textit{et al.} \cite{Dima} have carried out a detailed parametric study of two-electron temperature plasma sheath with nonextensive electrons and derived a modified sheath criterion incorporating the effect of nonextensive electrons. They have also shown that ion velocity at the sheath edge decreases with increasing ion-neutral collision \cite{Dima2}. Safa \textit{et at.} \cite{Safa2} have studied a magnetized plasma sheath with nonextensive electrons and showed that the modified Bohm velocity decreases with an increase in the nonextensive parameter $q$. Moulick \textit{et al.} \cite{Mou} extensively studied the combined effect of collision and nonextensive parameter $q$ emphasizing on the space charge deposition near a wall. 

Such theoretical investigations have firmly proved that the $q$ parameter plays a vital role in determining the ion dynamics inside the sheath.  In these studies, the $q$ parameter has been varied in the ranges $-1<q<1$ and $q>1$. But, are these models valid for all the allowed values of $q$? In other words, what is the most suitable value of $q$ for a specific plasma condition? A review of the literature reveals that theoretical/numerical investigations are silent on this particular issue. However, experimental studies have been able to predict the value of $q$ in different circumstances \cite{Bog, Du}. Qiu \textit{et al.} have initiated a measurement of the nonextensive parameter in laboratory plasmas proposing a nonextensive single electric probe \cite{Qiu}. They have modified the probe current expressions based on the nonextensive distribution function and experimentally predicted a value of $q=0.775$ for a low-temperature unmagnetized plasma. Therefore, it is understood from their study that any admissible value of $q$ can not be used for a selective plasma model. It will be tactical to predict the value of $q$ for a given plasma condition and such an attempt has been made in the present study. 

In the first place, a comparative study between a two-fluid model and a single fluid model considering the nonextensive distribution for electrons has been carried out. In the two-fluid model, the effect of electron neutral collision and magnetic field on the electrons have been taken into consideration along with the pressure gradient force and the sheath electric field. Since, like particle collisions do not contribute to total momentum change, hence only electron-neutral collision has been considered here. Comparing the results of these two models, a suitable range of $q$ has been determined. Now, the effect of an additional electron group having energy higher than the primary electron group has been investigated. A single fluid approach has been adopted and both the electron groups are described by the nonextensive distribution, where, the previously determined range of the nonextensive parameter $q$ is employed. 

\section{Theoretical model}\label{theo}
\begin{figure}
    \centering
    \includegraphics[width=0.4\textwidth]{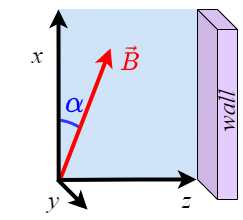}
    \caption{Schematic diagram of the  model}
    \label{fig: fig1}
\end{figure}

A collisional magnetized plasma is considered near a wall. The constant and uniform magnetic field is inclined to the wall in the $x-z$ plane making an angle $\alpha$ with the $x$-axis as shown in figure \ref{fig: fig1}.  A 1D3V fluid approach is adopted to solve the problem in hand. The continuity and momentum equations for ions are given as:
\begin{equation}\label{conti}
    \frac{d(n_iv_{iz})}{dz} = Zn_e,
\end{equation}

\begin{equation}\label{momi}
    n_i m_i v_{iz}\frac{d\textbf{v}_i}{dz} = n_i e\textbf{E} + n_i e(\textbf{v}_i \times \textbf{B}) - \nu_i m_i n_i\textbf{v}_i, 
\end{equation}

where, $n_i$ is the ion density, $n_e$ is the electron density, $Z$ is the ionization frequency and $\nu_{i}$ is the ion-neutral collision frequency. $\textbf{E}=-\frac{d\phi}{dz}\hat{k}$ is the electric field, $\phi$ is the electric potential, $m_i$ is the ion mass, and $\textbf{v}_i = v_{ix}\hat{i} + v_{iy}\hat{j} + v_{iz}\hat{k}$ is the ion fluid velocity.

The electrons are first described by the fluid equations without neglecting the inertia term. The corresponding continuity and momentum equations for the electrons are:
\begin{equation}\label{conte}
    \frac{d(n_ev_{ez})}{dz} = Zn_e,
\end{equation}

\begin{equation}\label{mome}
    n_e m_e v_{ez}\frac{d\textbf{v}_e}{dz} = -n_e e\textbf{E} - n_e e(\textbf{v}_e \times \textbf{B}) - \nu_e m_e n_e\textbf{v}_e - \frac{dp_e}{dz}\hat{k}, 
\end{equation}
where, $m_e$ is the mass of electron, $\nu_e$ is the electron neutral collision frequency, $T_e$ is the electron temperature, thermal pressure $~p_e=n_eT_e$ and $\textbf{v}_e = v_{ex}\hat{i} + v_{ey}\hat{j} + v_{ez}\hat{k}$ is the electron fluid velocity. The Boltzmann constant $k_B$ is taken inside $T_e$ throughout the paper.  These four equations are closed by the Poisson's equation
\begin{equation}\label{poi}
    \frac{d^2\phi}{dz^2} = -\frac{e}{\epsilon_0}(n_i-n_e),
\end{equation}
where, $\epsilon_0$ is the permitivity of free space. These five equations have constituted the two fluid model (M1). Results of this model is compared with another model where electron density is described by the nonextensive distribution as

\begin{equation}\label{e_den}
   n_e = n_{e0}\left(1+(q-1)\frac{e\phi}{T_e}\right)^{\frac{q+1}{2(q-1)}},
\end{equation}
where, $q$ is the nonextensive parameter. Equation \ref{conti}, \ref{momi}, \ref{poi} and \ref{e_den} constitutes the single fluid model with nonextensive electron distribution (M2).

The following dimensionless quantities are used to normalize the above set of equations.
$$
\begin{array}{cc}
\xi=\frac{z}{\lambda_{ni}},~~~u_i=\frac{v_{ix}}{c_s},~~~v_i=\frac{v_{iy}}{c_s},~~~w_i=\frac{v_{iz}}{c_s},\\~\\

u_e=\frac{v_{ex}}{c_s},~~~v_e=\frac{v_{ey}}{c_s},~~~w_e=\frac{v_{ez}}{c_s},\\~\\

\lambda_{ni}=\frac{c_s}{Z},~~~N_j=\frac{n_j}{n_{i0}},~~~\eta=-\frac{e\phi}{T_e},~~~\mu=\frac{m_i}{m_e},\\~\\
\gamma_{ik}=\frac{\lambda_{ni}}{c_s}\omega_{ik},~~~\gamma_{ek}=\frac{\lambda_{ni}}{c_s}\omega_{ek},~~~K_j=\frac{\lambda_{ni}}{c_s}\nu_j,~~~a_0=\frac{\lambda_{Di}}{\lambda_{ni}}.
\end{array}
$$
Here, $\lambda_{Di}$ is the ion Debye length, $\lambda_{ni}$ is the ionisation length, $c_s = \sqrt{T_e/m_i}$ is the ion sound speed, $\omega_{ik}$ and $\omega_{ek}$ are ion and electron gyro-frequency, $j=i,e$ and $k=x,z$.

The normalized forms of equation (\ref{conti})-(\ref{e_den}) are read as

\begin{equation}\label{ni}
    \frac{dN_i}{d\xi} = -\frac{N_i}{w_i^2}\left(\frac{d\eta}{d\xi}\right) + \gamma_{ix}N_i\left(\frac{v_i}{w_i^2}\right) + K_i\left(\frac{N_i}{w_i}\right) +  \left(\frac{N_e}{w_i}\right),
\end{equation}

\begin{equation}\label{ui}
    \frac{du_i}{d\xi} = \gamma_{iz} \left( \frac{v_i}{w_i} \right) - K_i\left(\frac{u_i}{w_i}\right),
\end{equation}

\begin{equation}\label{vi}
    \frac{dv_i}{d\xi} = \gamma_{ix} - \gamma_{iz} \left( \frac{u_i}{w_i} \right) - K_i \left( \frac{v_i}{w_i} \right),
\end{equation}

\begin{equation}\label{wi}
    \frac{dw_i}{d\xi} = \frac{1}{w_i} \left( \frac{d\eta}{d\xi} \right) - \gamma_{ix} \left( \frac{v_i}{w_i} \right) -K_i,
\end{equation}


\begin{equation}\label{ne}
    \frac{dN_e}{d\xi} = \left(\frac{w_e^2}{w_e^2-\mu}\right)\left( \frac{N_e}{w_e} + \frac{\mu N_e}{w_e^2}\left(\frac{d\eta}{d\xi}\right) + \gamma_{ex}\frac{N_ev_e}{w_e^2}\right),
\end{equation}

\begin{equation}\label{ue}
    \frac{du_e}{d\xi} = -\gamma_{ez} \left( \frac{v_e}{w_e} \right) - K_e\left(\frac{u_e}{w_e}\right),
\end{equation}

\begin{equation}\label{ve}
    \frac{dv_e}{d\xi} = -\gamma_{ex} + \gamma_{ez} \left( \frac{u_e}{w_e} \right) - K_e \left( \frac{v_e}{w_e} \right),
\end{equation}

\begin{equation}\label{we}
    \frac{dw_e}{d\xi} = \left(\frac{w_e^2}{w_e^2-\mu}\right) \left(-\frac{\mu}{w_e}\left(\frac{d\eta}{d\xi}\right) + \gamma_{ex}\left(\frac{v_e}{w_e}\right) - K_e + \frac{\mu}{w_e^2}  \right),
\end{equation}

\begin{equation}\label{po}
    \frac{d^2\eta}{d\xi^2} = \frac{1}{a_0^2}\left(N_i-N_e\right),
\end{equation}

\begin{equation}
    N_e = \left(1-(q-1)\eta\right)^{\frac{q+1}{2(q-1)}}.
\end{equation}

The parameters $K_i$ and $K_e$ appeared in the normalized equations are the ratio of ion-neutral collision frequency to ionisation frequency and electron-neutral collision frequency to ionisation frequency respectively. These two parameters can be treated as measures of plasma collisionality. Here, to model plasma collisionality, the constant collision frequency model has been adopted \cite{kishor, maso}.
\section{Numerical Execution}
The point $\xi=0$ is considered as the presheath boundary from where the numerical integration is started. The normalized bulk ion (electron) density is used as initial ion (electron) density. The initial value of species velocities and plasma potential are estimated near this boundary by employing the Taylor series solution method. The following series are used for this purpose \cite{RM}. 

\[
     \textbf{v}_k= \sum_s\textbf{v}_{ks}\xi^{2n+1}
\]
\[
     \eta = \sum_s\eta_s\xi^{2n}
\]

The mentioned series are used in the governing equations and the first order Taylor co-efficients obtained after series expansion are treated as initial values. The standard Matlab routine ODE45 has been employed to solve the described set of equations. The following default parameters are used keeping the view that results of this study might find applications in low pressure gas discharges.

$$\begin{array}{cc}
n_{i0}=10^{16}~m^{-3},~~~T_e=2.0~eV,~~~T_i=0.026~eV,~~~Z=10^5~s^{-1}.
\end{array}$$

\section{Results and Discussion}

\begin{figure}[h!]
    \centering
    \begin{minipage}[b]{0.48\textwidth}
    \includegraphics[width=1\textwidth]{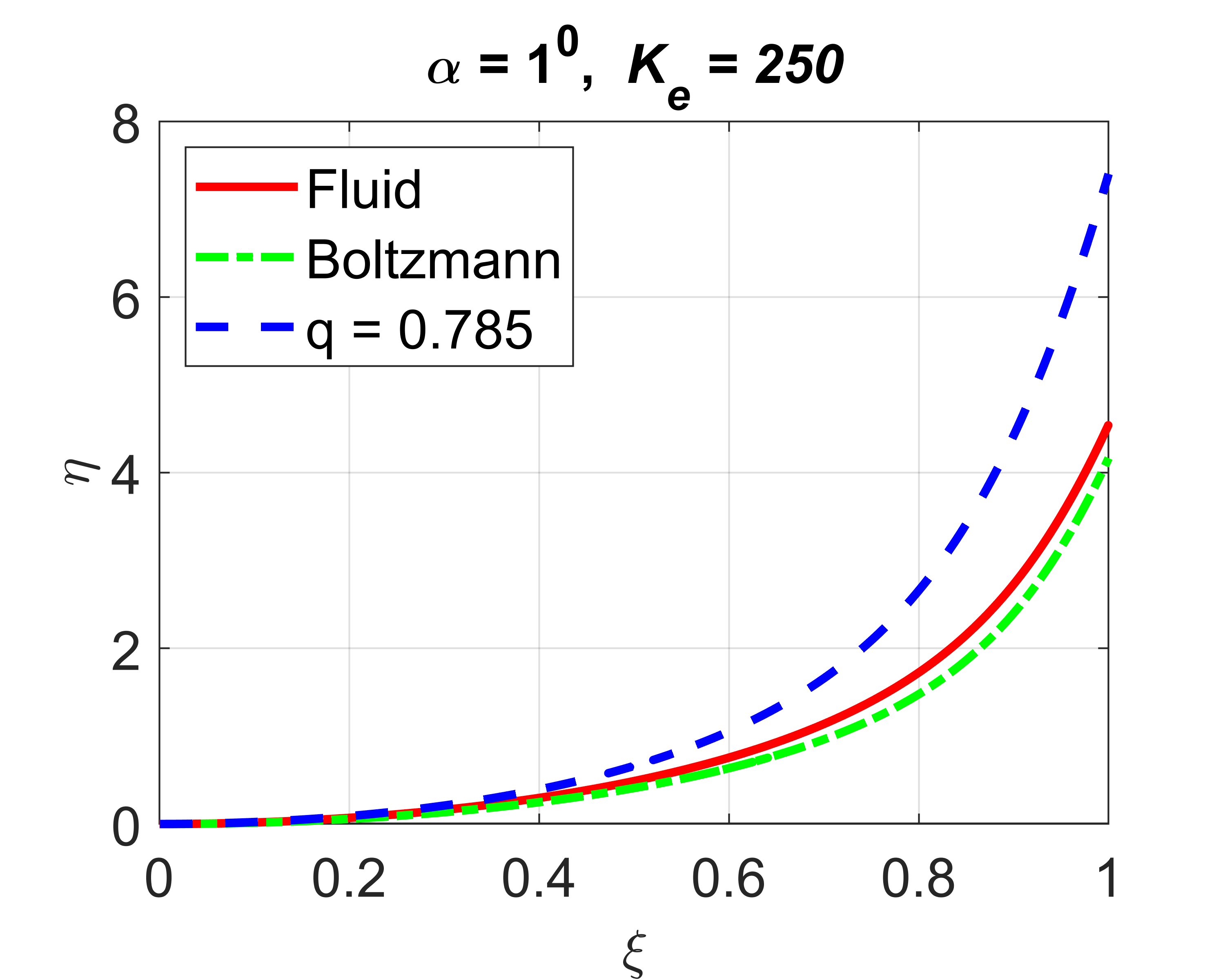}
    \caption{Distribution of normalized sheath potential for different electron models}
    \label{fig:2}
    \end{minipage}
    \begin{minipage}[b]{0.48\textwidth}
    \includegraphics[width=1\textwidth]{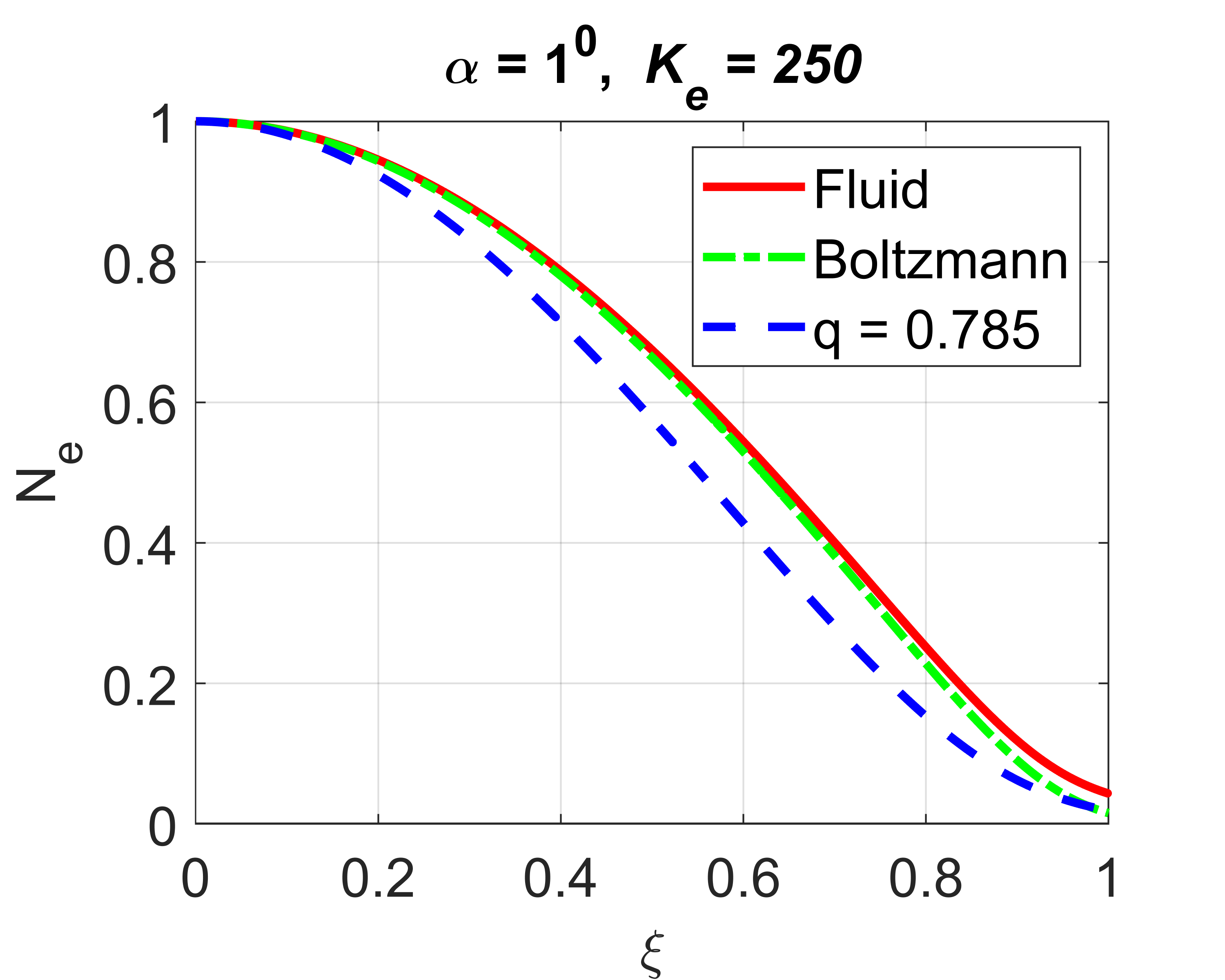}
    \caption{Distribution of normalized electron density for different electron models}
    \label{fig:3}
    \end{minipage}
\end{figure}

\begin{figure}[h!]
    \centering
    \begin{minipage}[b]{0.48\textwidth}
    \includegraphics[width=1\textwidth]{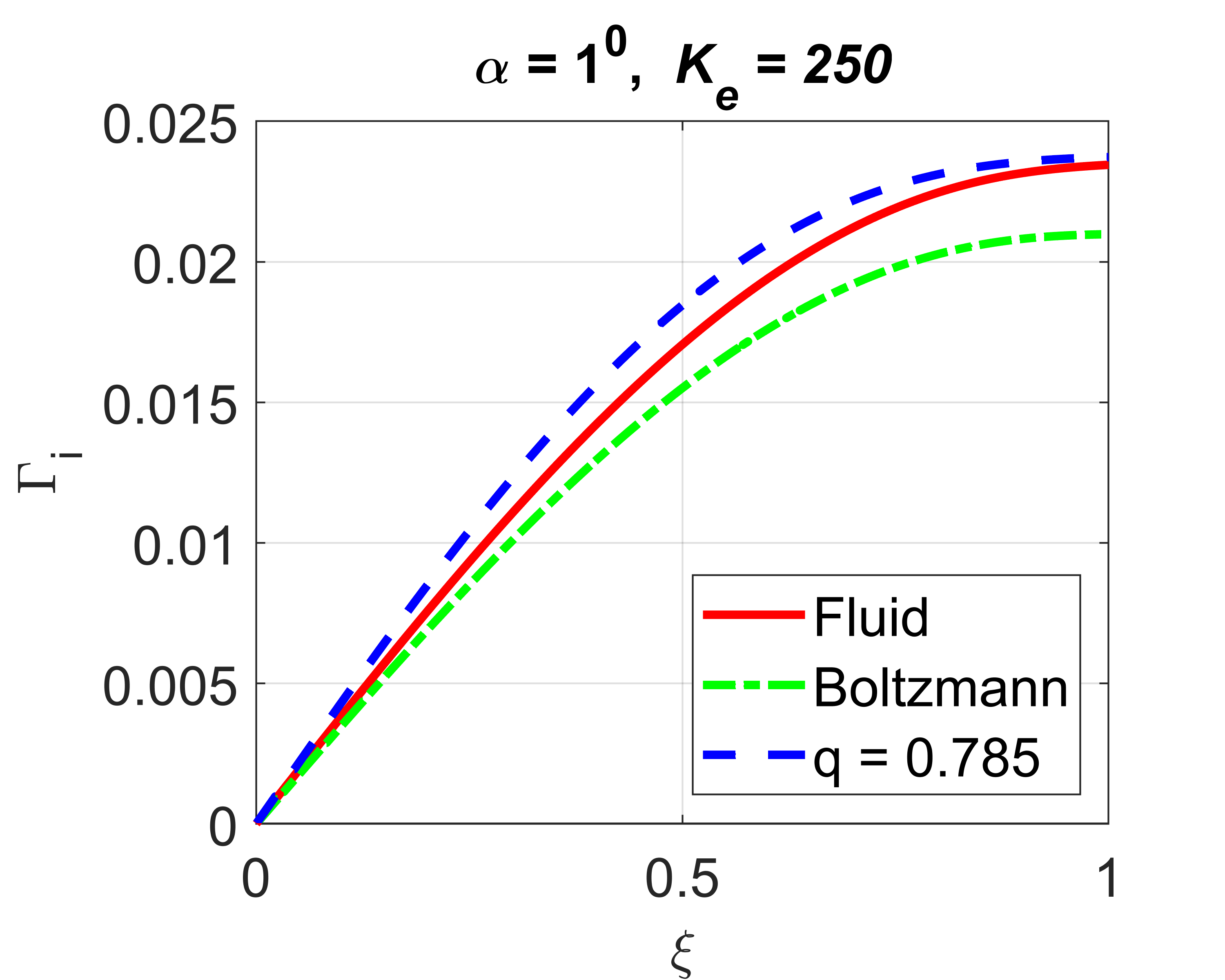}
    \caption{Variation of normalized ion flux in the sheath for different electron models}
    \label{fig:4}
    \end{minipage}
    \begin{minipage}[b]{0.48\textwidth}
    \includegraphics[width=1\textwidth]{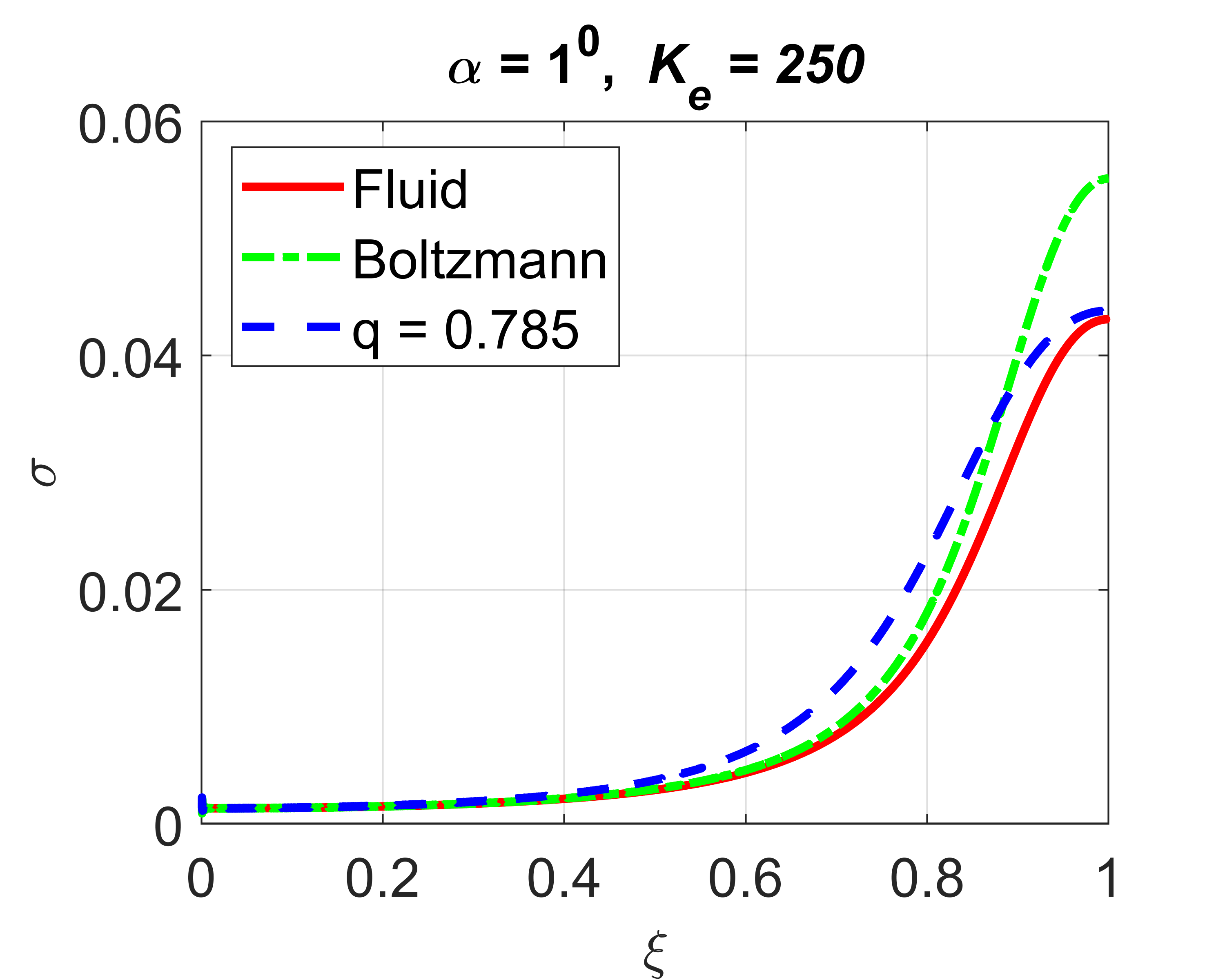}
    \caption{Variation of normalized space charge for different electron models}
    \label{fig:5}
    \end{minipage}
\end{figure}
\subsection{Comparative study between model M1 and model M2}
The validity of the Boltzmann distribution for electrons in the presence of an external magnetic field has been a matter of dispute for a long time \cite{RN}. The effect of an oblique magnetic field on the electrons is often ignored either on the basis of their small fluid velocity in comparison to their thermal velocity or by considering a weak magnetic field \cite{Chodu, Hat}. Another argument is equally proposed in support of the consideration where it is assumed that due to their high thermal velocity, electrons are strongly magnetized. Hence, their guiding centers follow the magnetic field, and the Boltzmann distribution is retained \cite{RM,PC}. But, if the angle of inclination of the magnetic field to the tangent of the wall is very small ($\alpha \rightarrow 0$), the Boltzmann distribution of electron is doubtful \cite{Tsa2}. Moreover, in the divertor region of tokamaks, the maximum inclination of the magnetic field towards the wall is $5^0$ \cite{PC}. Therefore, in this study, the inclination angle $\alpha$ is kept between $1^0\leq\alpha\leq5^0$ having magnitude 1 Tesla, which is a decisive range in fusion devices. 

Let us consider the low electron neutral collision regime. A comparison of electric potential, electron density, ion flux and space charge have been depicted in figure \ref{fig:2}, figure \ref{fig:3}, figure \ref{fig:4} and figure \ref{fig:5} respectively among various electron models. In all the plots, the deviation of the Boltzmann distributed electron model from the two-fluid model for an almost parallel magnetic field to the wall is evident. In figure \ref{fig:2}, for $q=0.785$, a comparatively higher sheath potential is observed. However, for the same value of $q$, the measured values of ion flux and the maximum space charge deposited near the wall are in reasonably good agreement with the two-fluid models. 

\begin{figure}[h!]
    \centering
    \begin{minipage}[b]{0.48\textwidth}
    \includegraphics[width=1\textwidth]{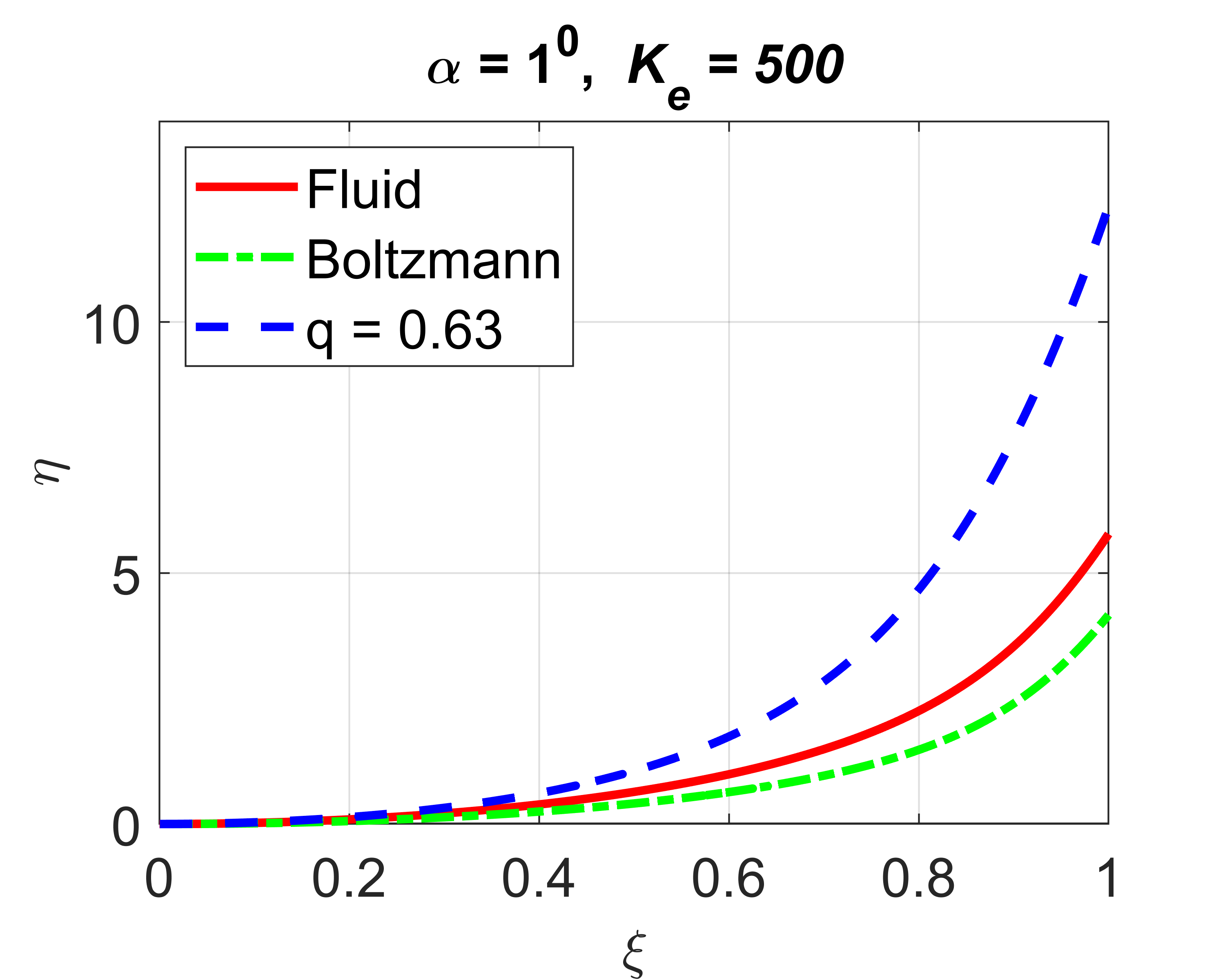}
    \caption{Distribution of normalized sheath potential for different electron models}
    \label{fig:6}
    \end{minipage}
    \begin{minipage}[b]{0.48\textwidth}
    \includegraphics[width=1\textwidth]{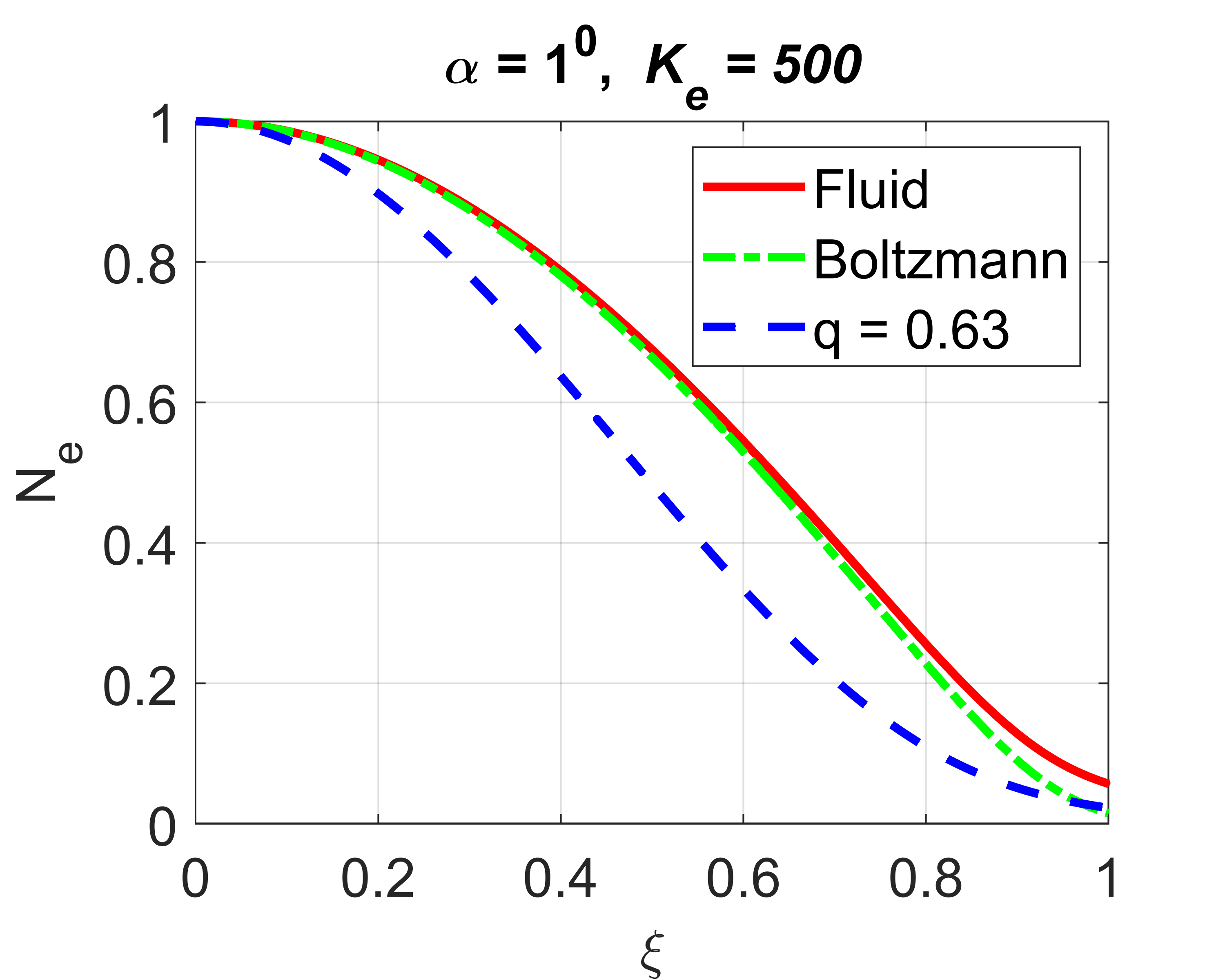}
    \caption{Distribution of normalized electron density for different electron models}
    \label{fig:7}
    \end{minipage}
\end{figure}

\begin{figure}[h!]
    \centering
    \begin{minipage}[b]{0.48\textwidth}
    \includegraphics[width=1\textwidth]{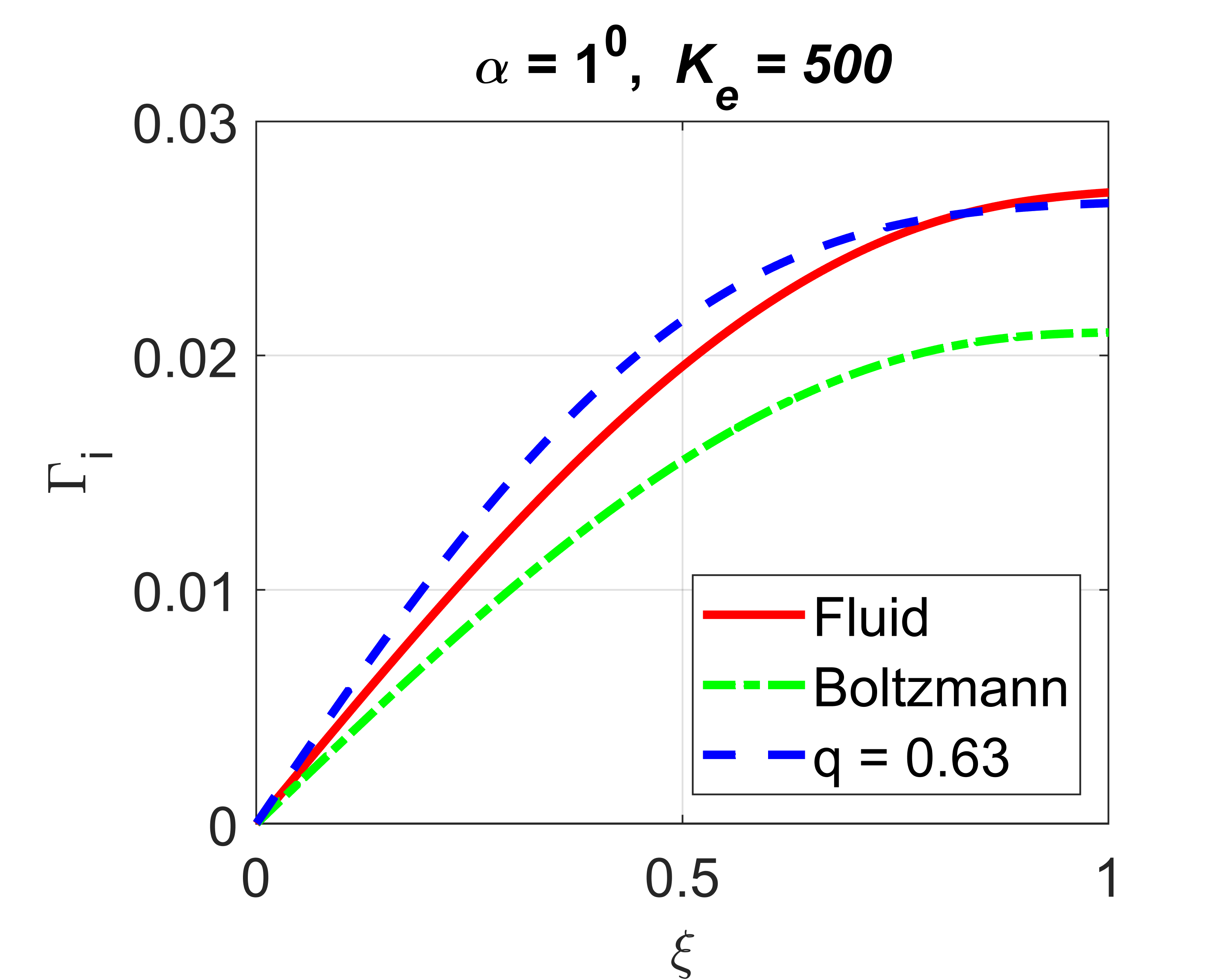}
    \caption{Variation of normalized ion flux  for different electron models}
    \label{fig:8}
    \end{minipage}
    \begin{minipage}[b]{0.48\textwidth}
    \includegraphics[width=1\textwidth]{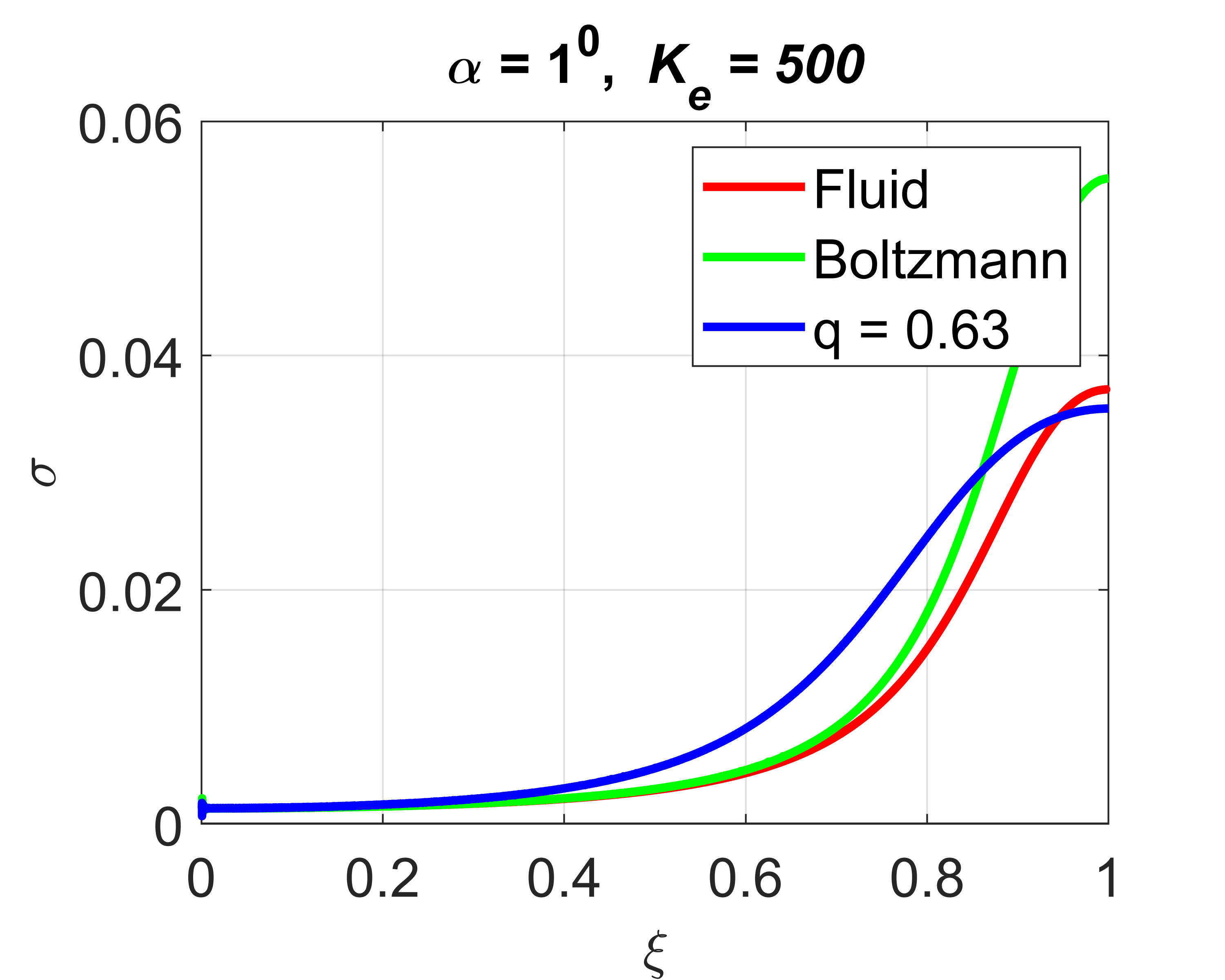}
    \caption{Variation of normalized space charge for different electron models}
    \label{fig:9}
    \end{minipage}
\end{figure}

Moving to a higher electron neutral collision regime, a similar set of comparative results are portrayed in figure \ref{fig:6}, \ref{fig:7}, \ref{fig:8} and figure \ref{fig:9}. It is observed that an increase in the electron collision parameter $K_e$ leads to a greater deviation of electron distribution from the usual Boltzmann distribution. For a value of $q=0.63$, the ion flux and space charge distribution become comparable to those obtained from the fluid model. But again, a higher electric potential in comparison to the fluid model has been observed. 

\begin{figure}
    \centering
    \begin{minipage}[b]{0.48\textwidth}
    \includegraphics[width=1\textwidth]{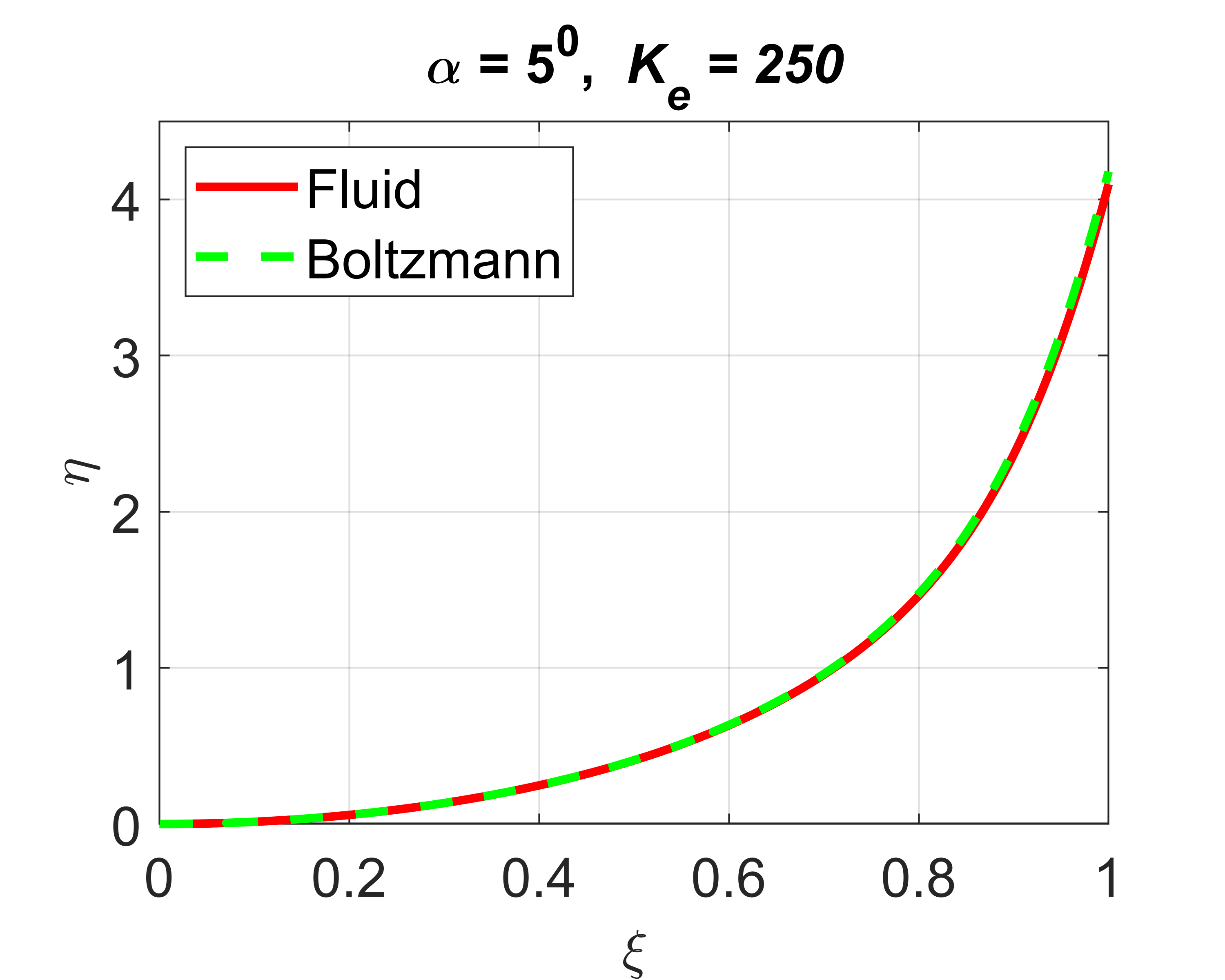}
    \caption{Variation of normalized electric potential in the sheath for fluid and Boltzmann electrons}
    \label{fig:10}
    \end{minipage}
    \begin{minipage}[b]{0.48\textwidth}
    \includegraphics[width=1\textwidth]{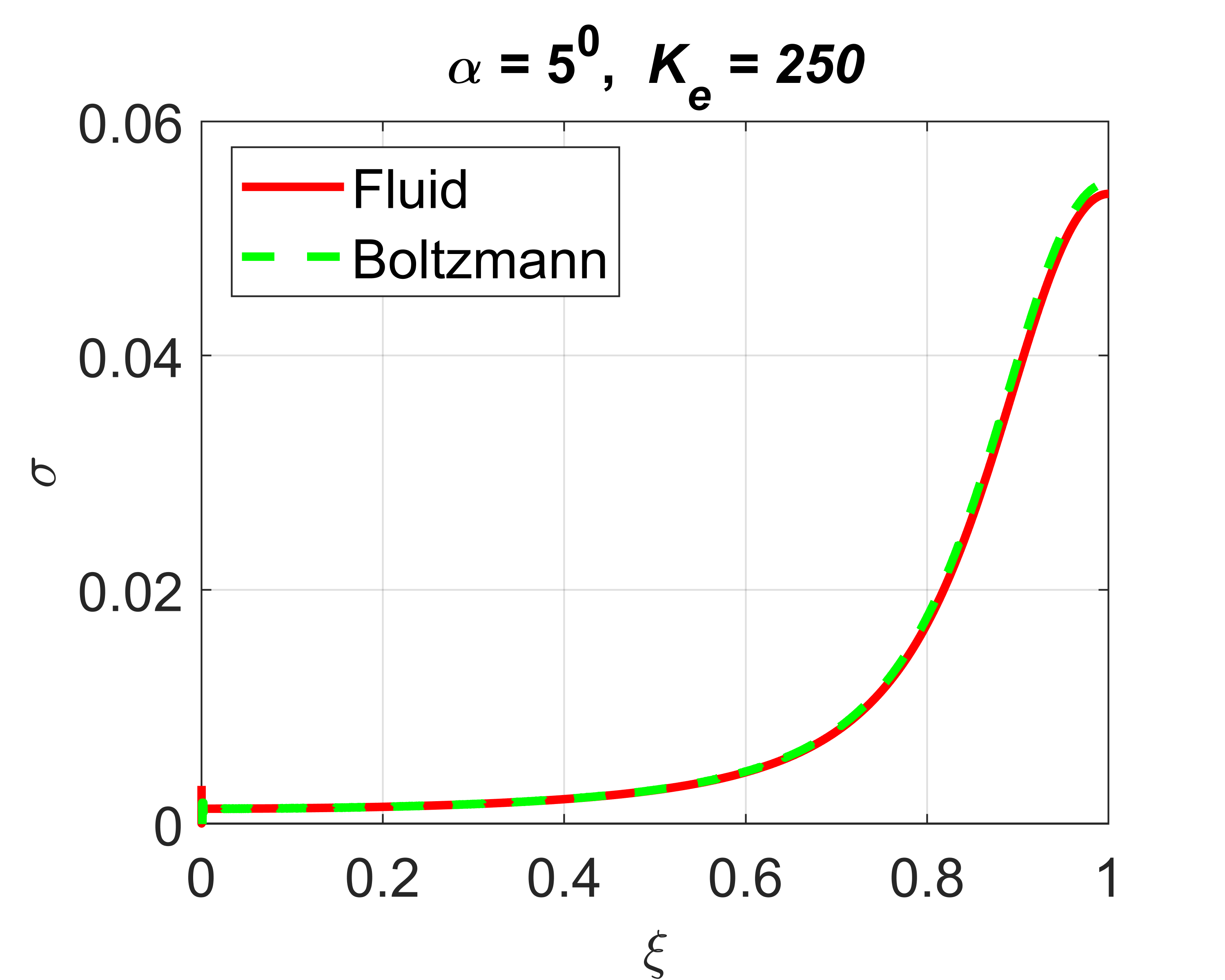}
    \caption{Variation of normalized space charge in the sheath for fluid and Boltzmann electrons}
    \label{fig:11}
    \end{minipage}
\end{figure}

\begin{figure}
    \centering
    \begin{minipage}[b]{0.48\textwidth}
    \includegraphics[width=1\textwidth]{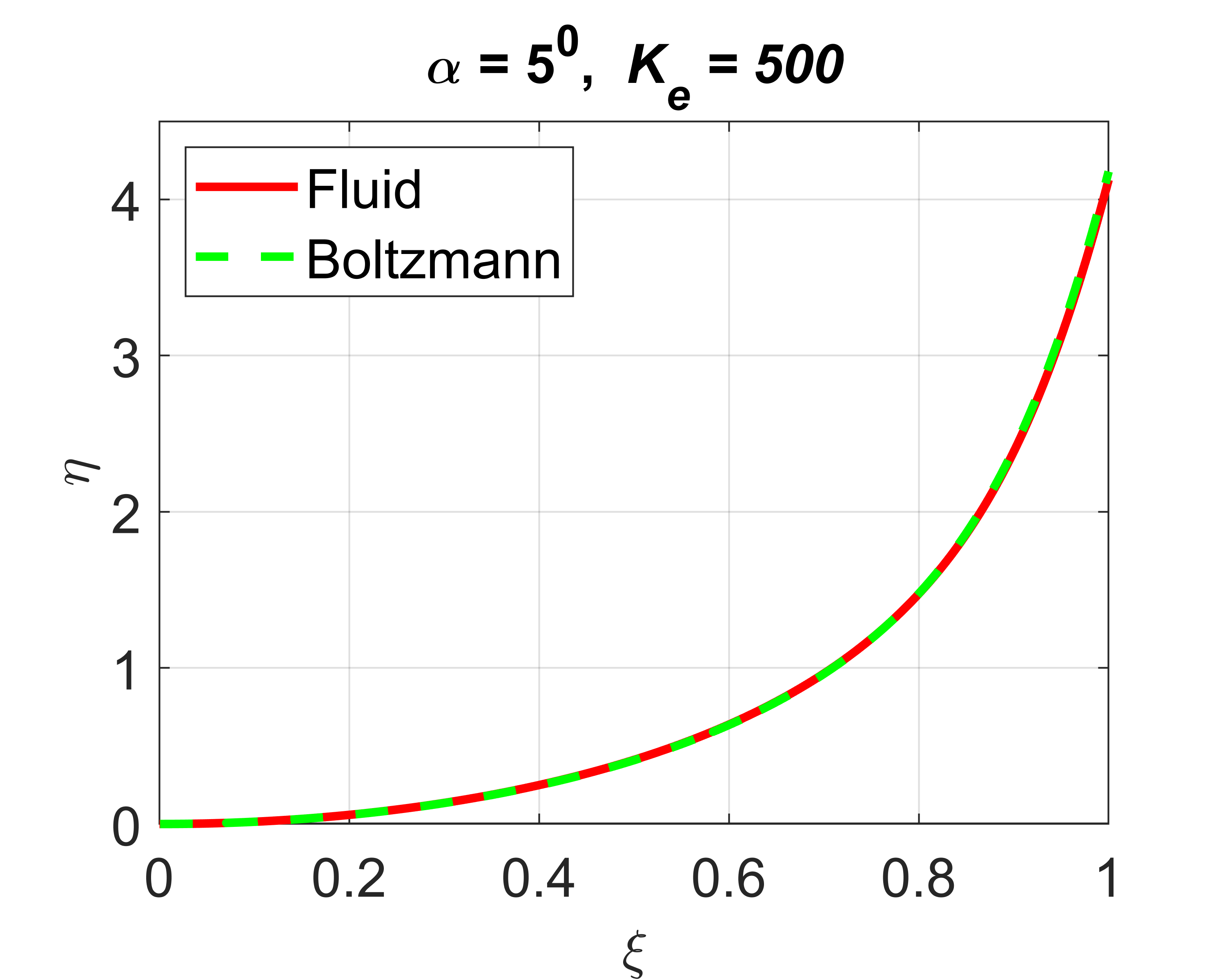}
    \caption{Variation of normalized electric potential in the sheath for fluid and Boltzmann electrons}
    \label{fig:12}
    \end{minipage}
    \begin{minipage}[b]{0.48\textwidth}
    \includegraphics[width=1\textwidth]{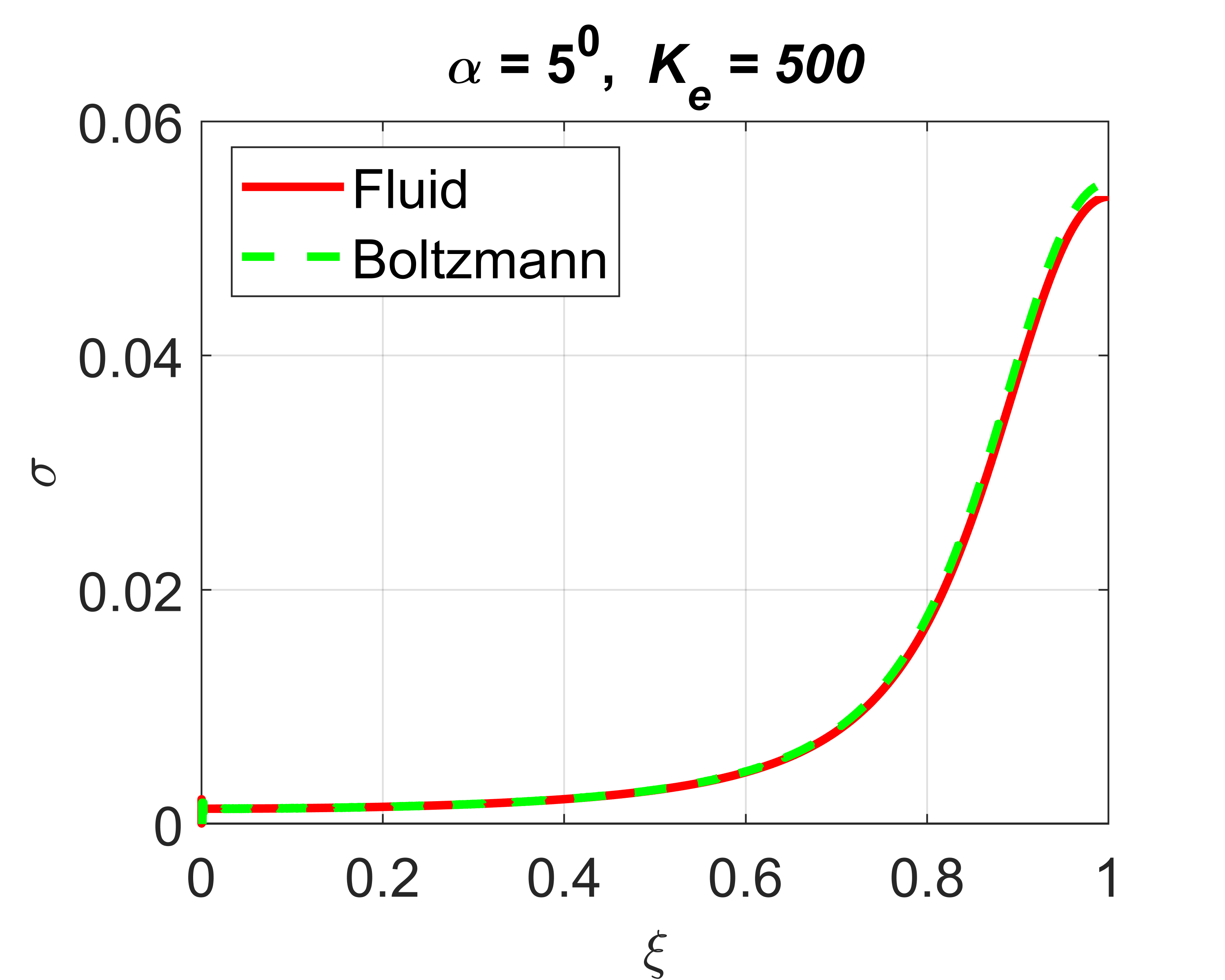}
    \caption{Variation of normalized space charge in the sheath for fluid and Boltzmann electrons}
    \label{fig:13}
    \end{minipage}
\end{figure}

\begin{figure}
    \centering
    \includegraphics[width=0.5\textwidth]{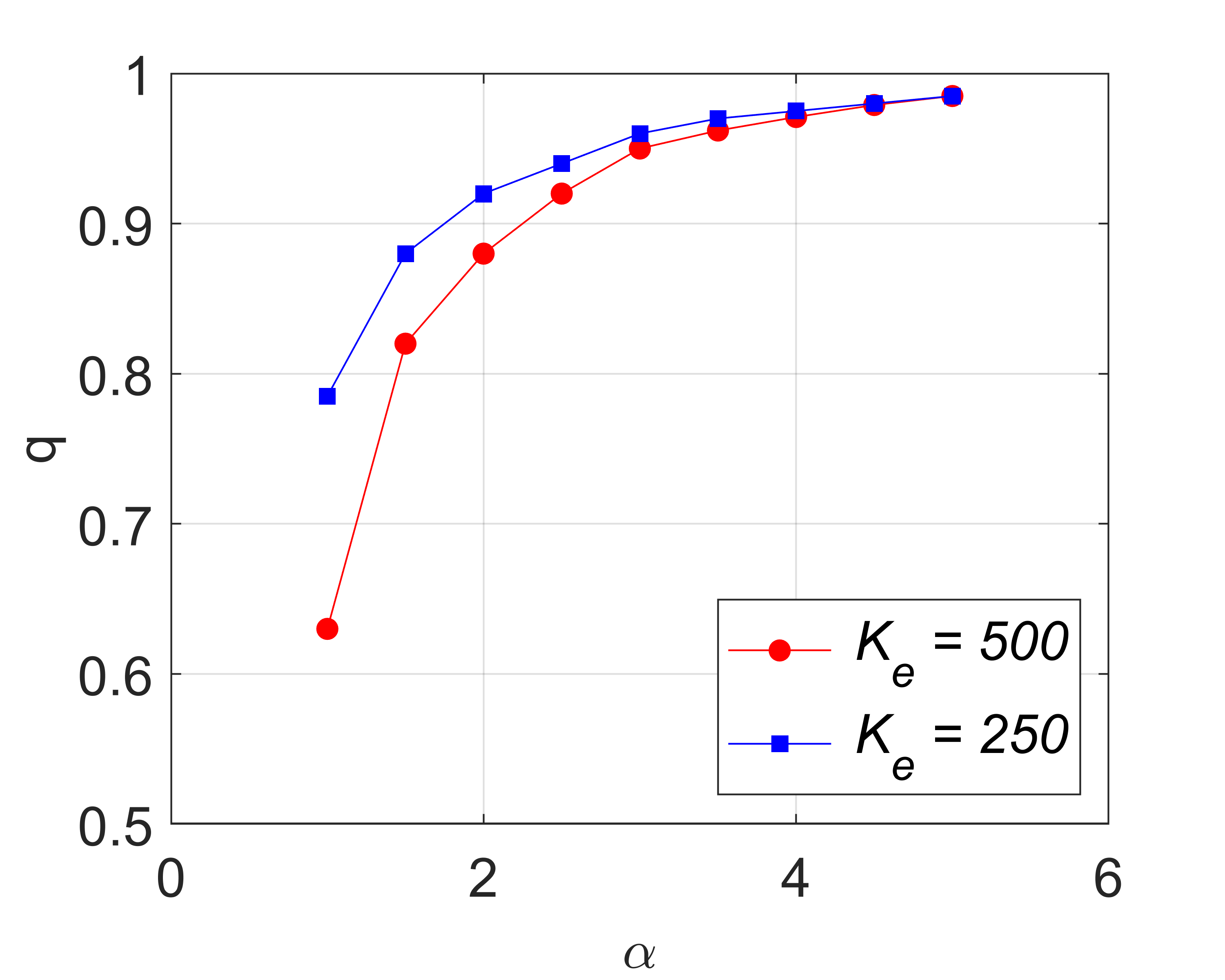}
    \caption{Variation of $q$ with $\alpha$ in different collisional environments}
    \label{fig:alpha_q}
\end{figure}

It is well known that Boltzmann distribution for electrons physically signifies the balance between electrostatic and pressure gradient force. In a system, where other force fields are also present, this simple balance between electrostatic and pressure gradient force no longer holds good. The results of the comparative study discussed also suggests the same. Therefore one can infer that for partially ionized plasmas having higher electron neutral collision frequency, the electrons are no longer Boltzmann distributed. In such cases, a suitable range of the nonextensive parameter could be $0.5\leq q \leq 1$. But, what happens for a slightly higher angle of inclination of the magnetic field is entirely different. For the low collisional regime, figure \ref{fig:10} and figure \ref{fig:11}, and for the higher collisional regime, figure \ref{fig:12} and figure \ref{fig:13} represents the electric potential and space charge in the sheath respectively where the magnetic field is inclining at an angle $\alpha=5^0$ towards the wall. The profiles of sheath potential and space charge obtained from fluid electrons are exactly similar to those obtained from Boltzmann electrons. The dependency of the $q$ parameter on the angle of inclination $\alpha$ has been portrayed in the figure \ref{fig:alpha_q} for two different collision conditions. This comparison affirms that unless the external magnetic field is nearly parallel to the wall, the Boltzmann distribution for electrons can be safely deployed.

\subsection{Study of nontextensive two electron temperature plasma}
The comparative study discussed in the previous section yields a suitable range for $q$ parameter for a partially ionized magnetized plasma. A two-electron temperature plasma system has now been taken into consideration and analysis of sheath characteristics has been carried out in the context of nonextensive distribution for electrons. Both low and high-temperature groups of electrons are described by the nonextensive distribution. The low-temperature electrons are termed as cold electrons and the high-temperature electrons are termed as hot electrons. Their respective densities are expressed as 

\begin{equation}\label{c_den}
   n_c = n_{c0}\left(1+(q-1)\frac{e\phi}{T_c}\right)^{\frac{q+1}{2(q-1)}},
\end{equation}

\begin{equation}\label{c_den}
   n_h = n_{h0}\left(1+(q-1)\frac{e\phi}{T_h}\right)^{\frac{q+1}{2(q-1)}},
\end{equation}
where, $n_{c0}$ and $n_{h0}$ are the cold and hot electron densities in the bulk plasma and $T_c$ and $T_h$ are the cold and hot electron temperatures. The corresponding normalized equations are given by:
\begin{equation}\label{Nc}
    N_c = (1-\delta)\left(1-(q-1)\eta\right)^{\frac{q+1}{2(q-1)}},
\end{equation}

\begin{equation}\label{Nh}
    N_h = \delta \left(1-(q-1)\eta/\tau\right)^{\frac{q+1}{2(q-1)}},
\end{equation}
where, $\delta = n_{h0}/n_{i0}$ and $\tau=T_h/T_c$. The set of equations (\ref{ni})-(\ref{wi}) along with equation (\ref{po}), equation (\ref{Nc}) and equation (\ref{Nh}) constitute the two-electron temperature plasma sheath model with nonextensive electron distribution. In this study, the parametric dependence of ion flux and sheath width on $\delta$, $\tau$ and $K_i$ has been investigated in the predetermined range of $q$.

\begin{figure}
    \centering
    \begin{minipage}[b]{0.48\textwidth}
    \includegraphics[width=1\textwidth]{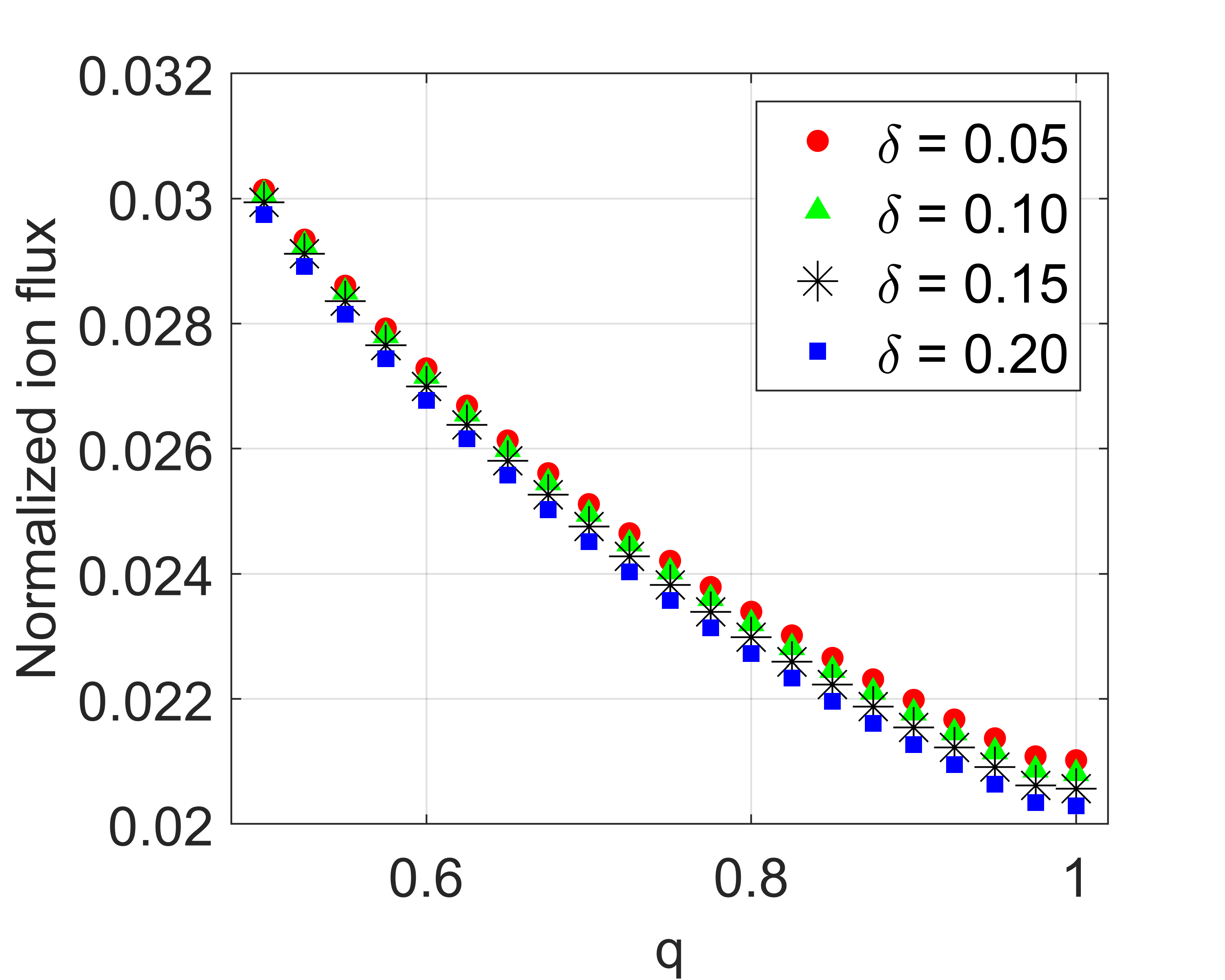}
    \caption{Variation of normalized ion flux with $q$ for different concentrations of hot electrons}
    \label{fig:14}
    \end{minipage}
    \begin{minipage}[b]{0.48\textwidth}
    \includegraphics[width=1\textwidth]{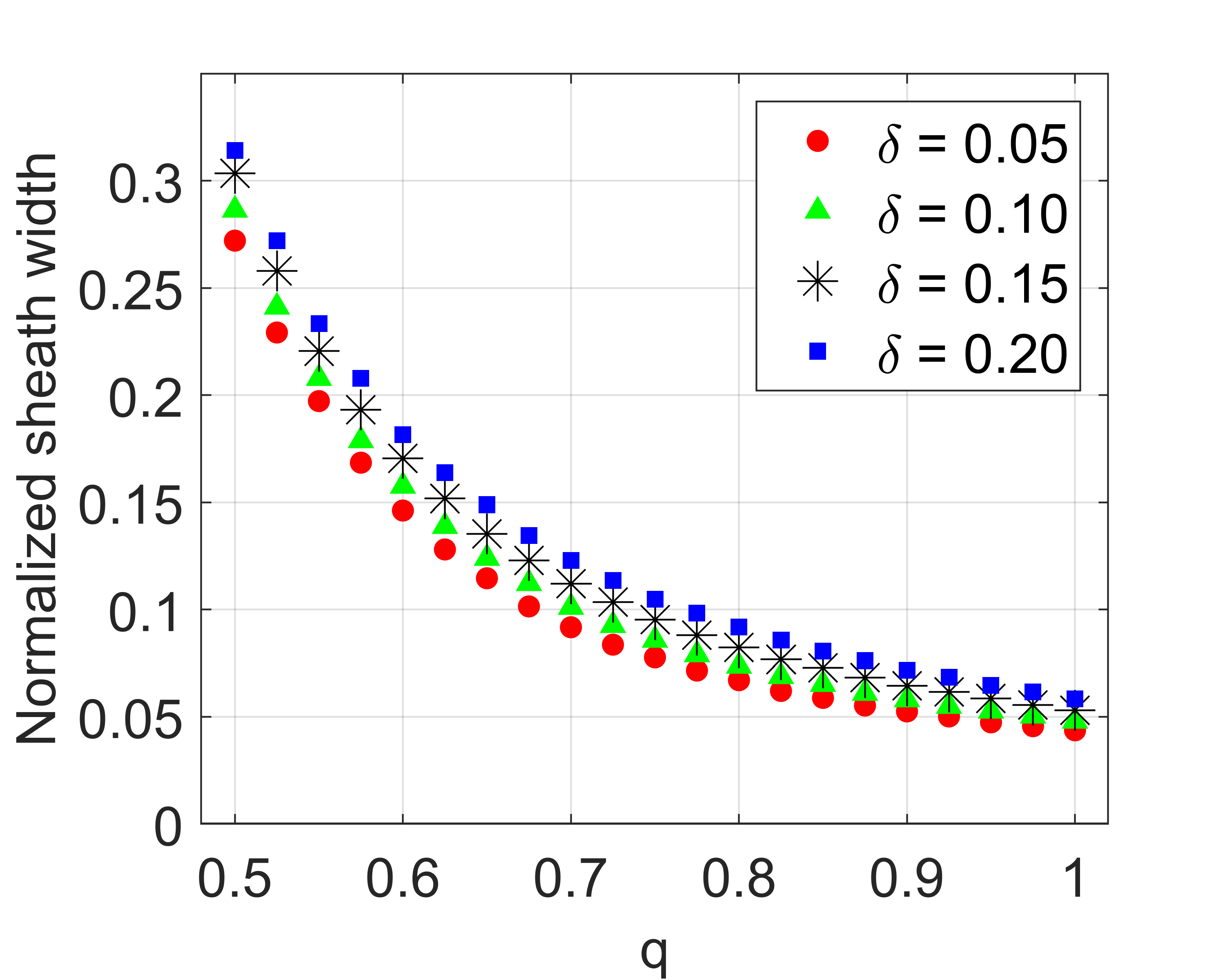}
    \caption{Variation of normalized sheath thickness with $q$ for different concentrations of hot electrons}
    \label{fig:15}
    \end{minipage}
\end{figure}

The behavior of ion flux and sheath thickness with the nonextensive parameter $q$ are depicted in figure \ref{fig:14} and figure \ref{fig:15} for four different values of $\delta$. In both cases, the value of the property in consideration decreases on moving towards $q=1$. The fall of electron density inside the sheath is steeper for the nonextensive distribution as compared to the Boltzmann distribution. As a result, a high potential drop is recorded for $q<1$ (Refer to figure (\ref{fig:2}) and figure (\ref{fig:6})). This increases the electric field in the region thereby enhancing the ion velocity. Hence, with decreasing $q$, the ion flux increases.
Again, a longer distance is required to shield a higher electric field. Therefore, the sheath formed in front of the wall also expands as electron distribution deviates from the Boltzmann distribution. On the other hand, ion flux calculated at the wall subsides with an increase in hot electron density. Hot electrons have enough thermal energy to overcome the sheath potential barrier. As a result, the total electron density in the sheath increases. This eventually limits the growth of space charge near the wall and reduces the sheath electric field and consequently, the ion flux decreases. 
On the other hand, the presence of hot electrons has an interesting effect on sheath thickness. A low electric field should have been shielded in a shorter length span but the overall particle density inside the sheath grows up with the raise in $\delta$ as stated above. Therefore, the sheath thickness is found to increases with the increase in hot electron population.

\begin{figure}
    \centering
    \begin{minipage}[b]{0.48\textwidth}
    \includegraphics[width=1\textwidth]{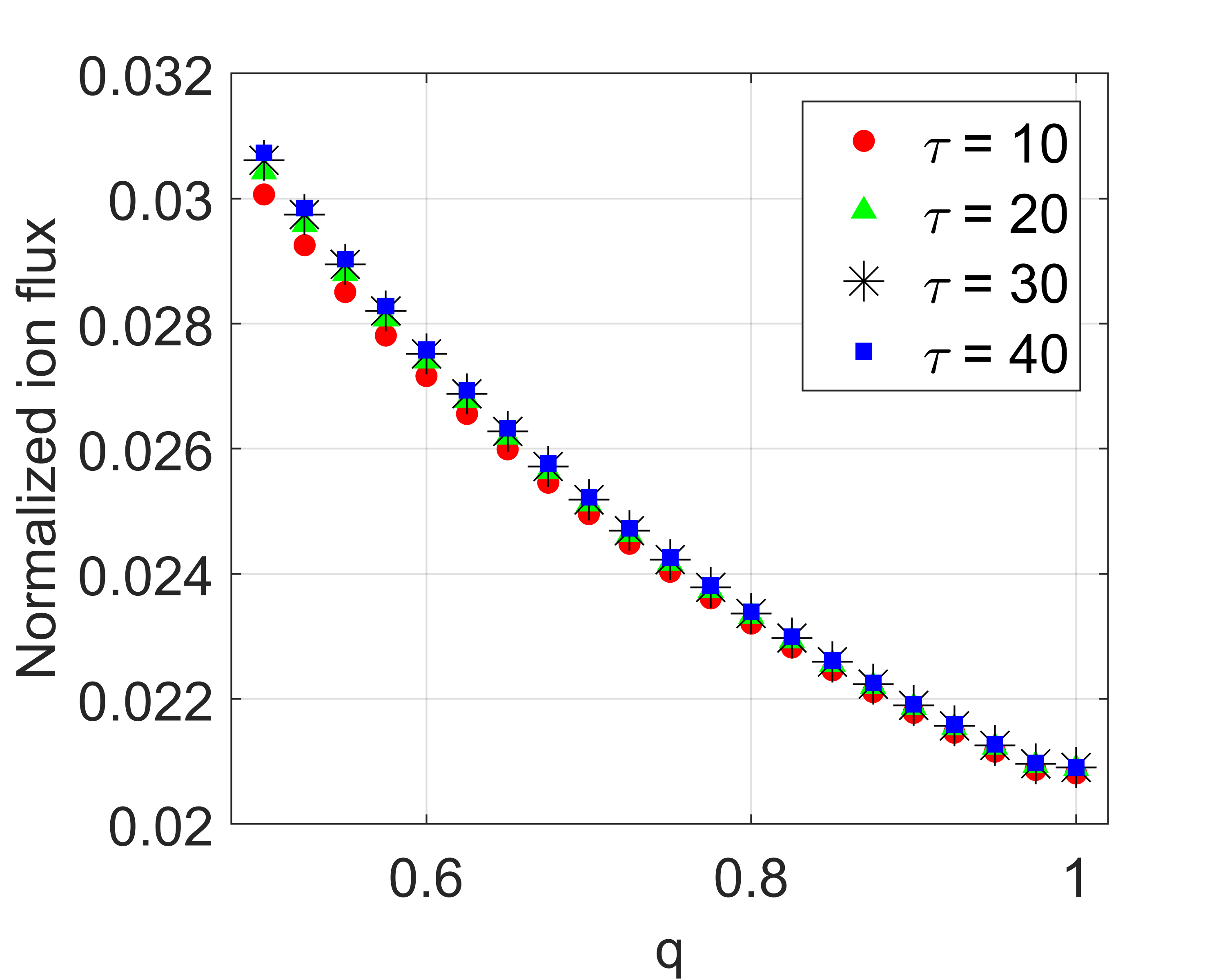}
    \caption{Variation of normalized ion flux with $q$ for different temperatures of hot electrons}
    \label{fig:16}
    \end{minipage}
    \begin{minipage}[b]{0.48\textwidth}
    \includegraphics[width=1\textwidth]{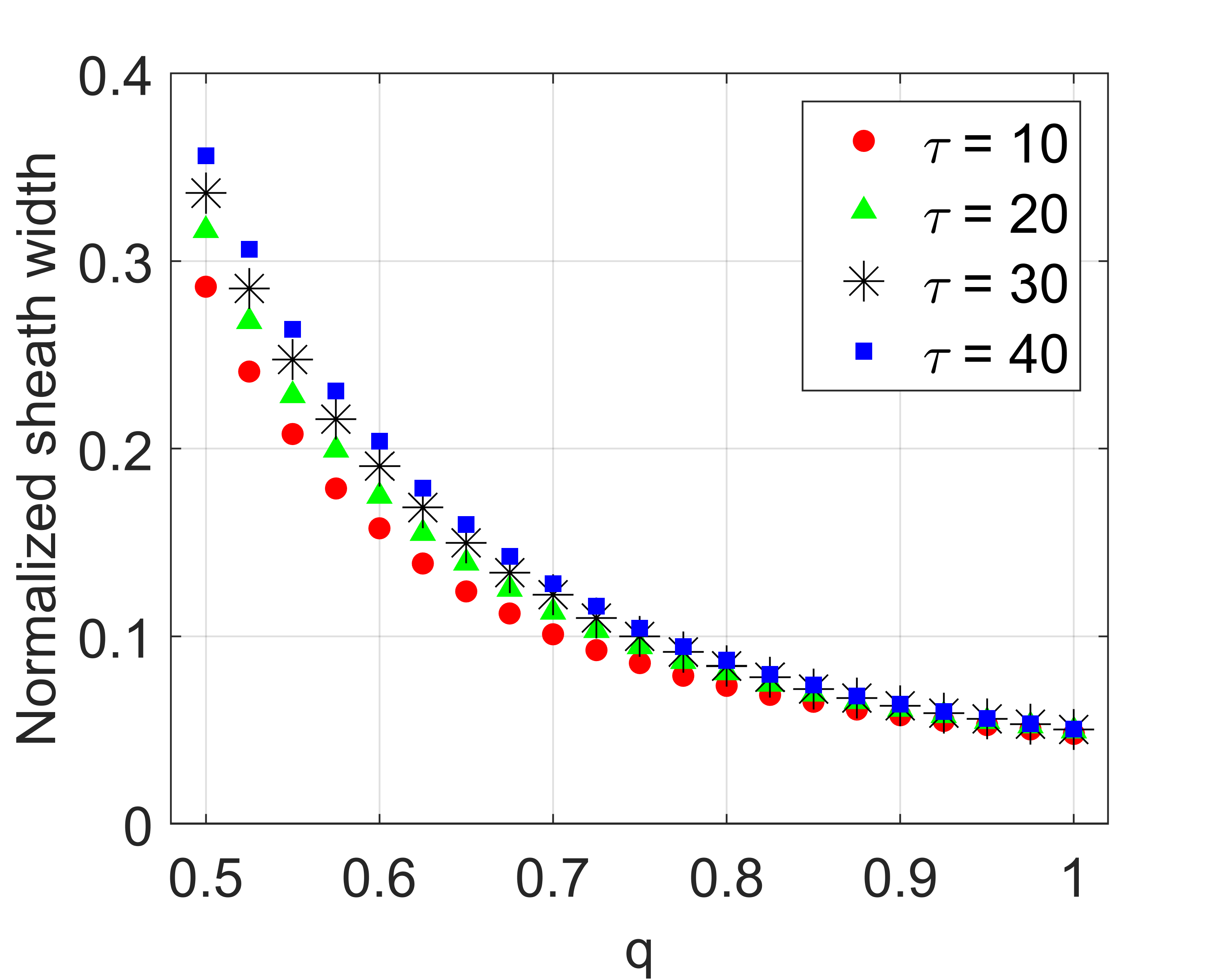}
    \caption{Variation of normalized sheath thickness with $q$ for different temperatures of hot electrons}
    \label{fig:17}
    \end{minipage}
\end{figure}
\begin{figure}
    \centering
    \begin{minipage}[b]{0.48\textwidth}
    \includegraphics[width=1\textwidth]{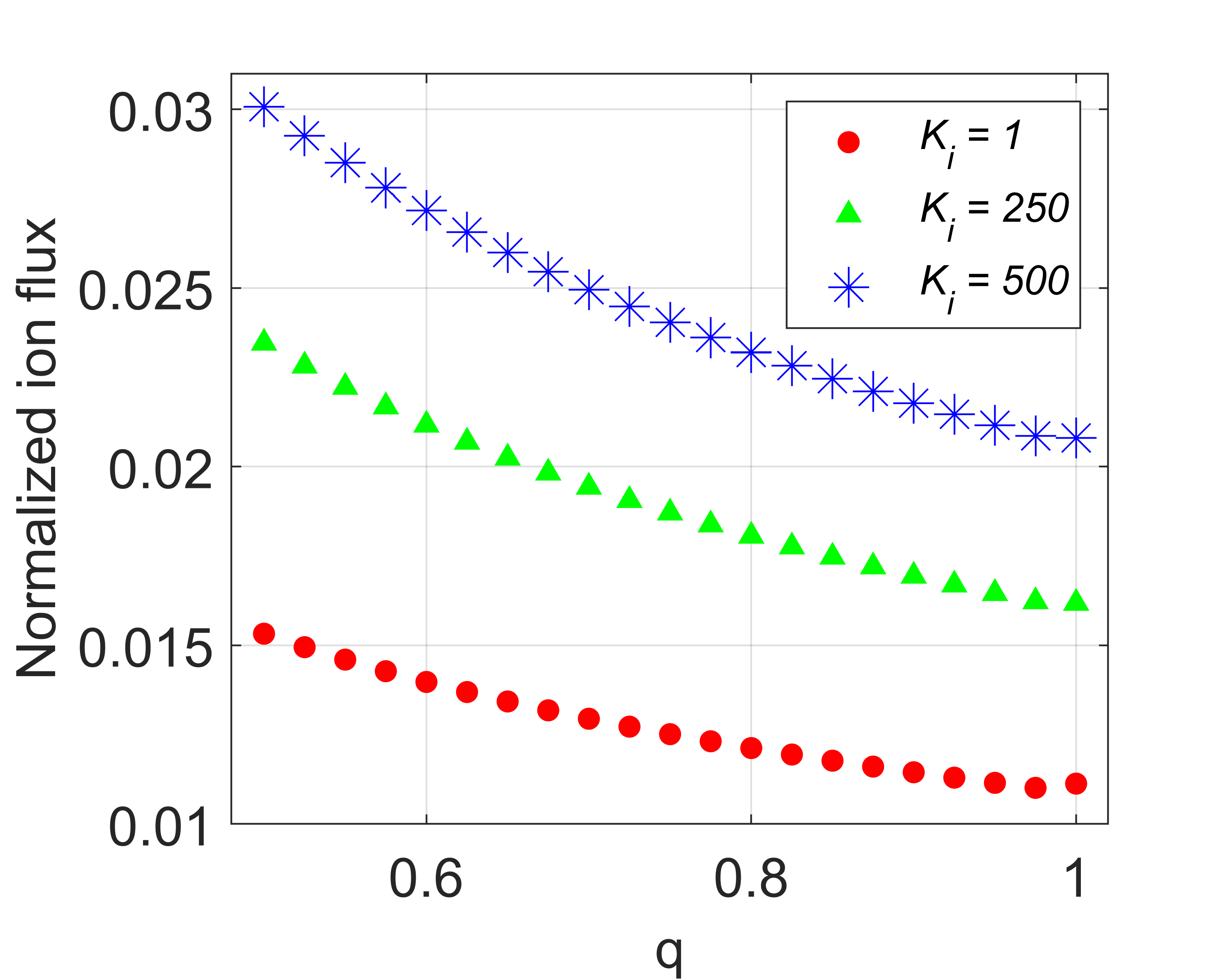}
    \caption{Variation of normalized ion flux with $q$ for various $K_i$}
    \label{fig:18}
    \end{minipage}
    \begin{minipage}[b]{0.48\textwidth}
    \includegraphics[width=1\textwidth]{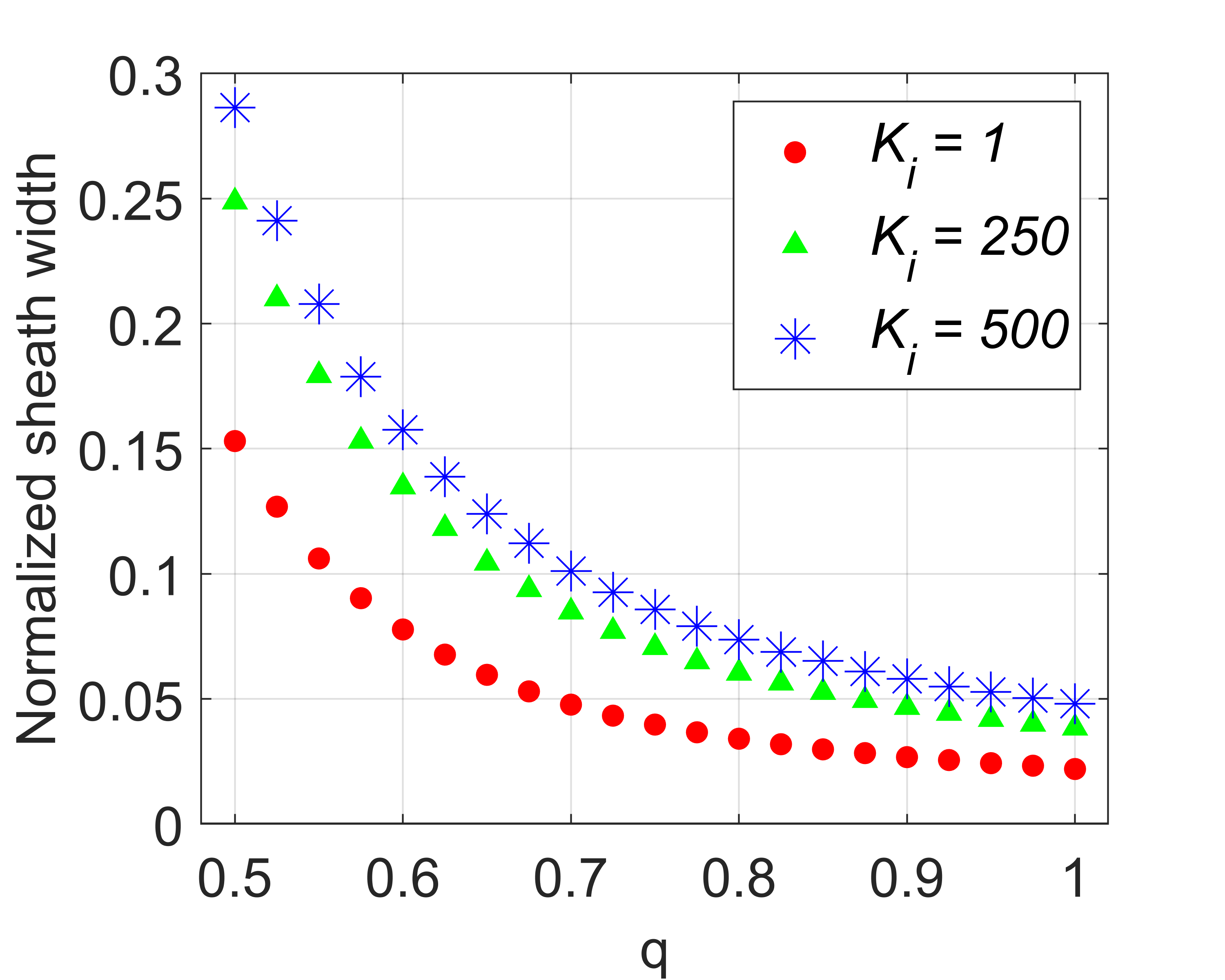}
    \caption{Variation of sheath thickness with $q$ for different $K_i$}
    \label{fig:19}
    \end{minipage}
\end{figure}
\begin{figure}
    \centering
    \begin{minipage}[b]{0.48\textwidth}
    \includegraphics[width=1\textwidth]{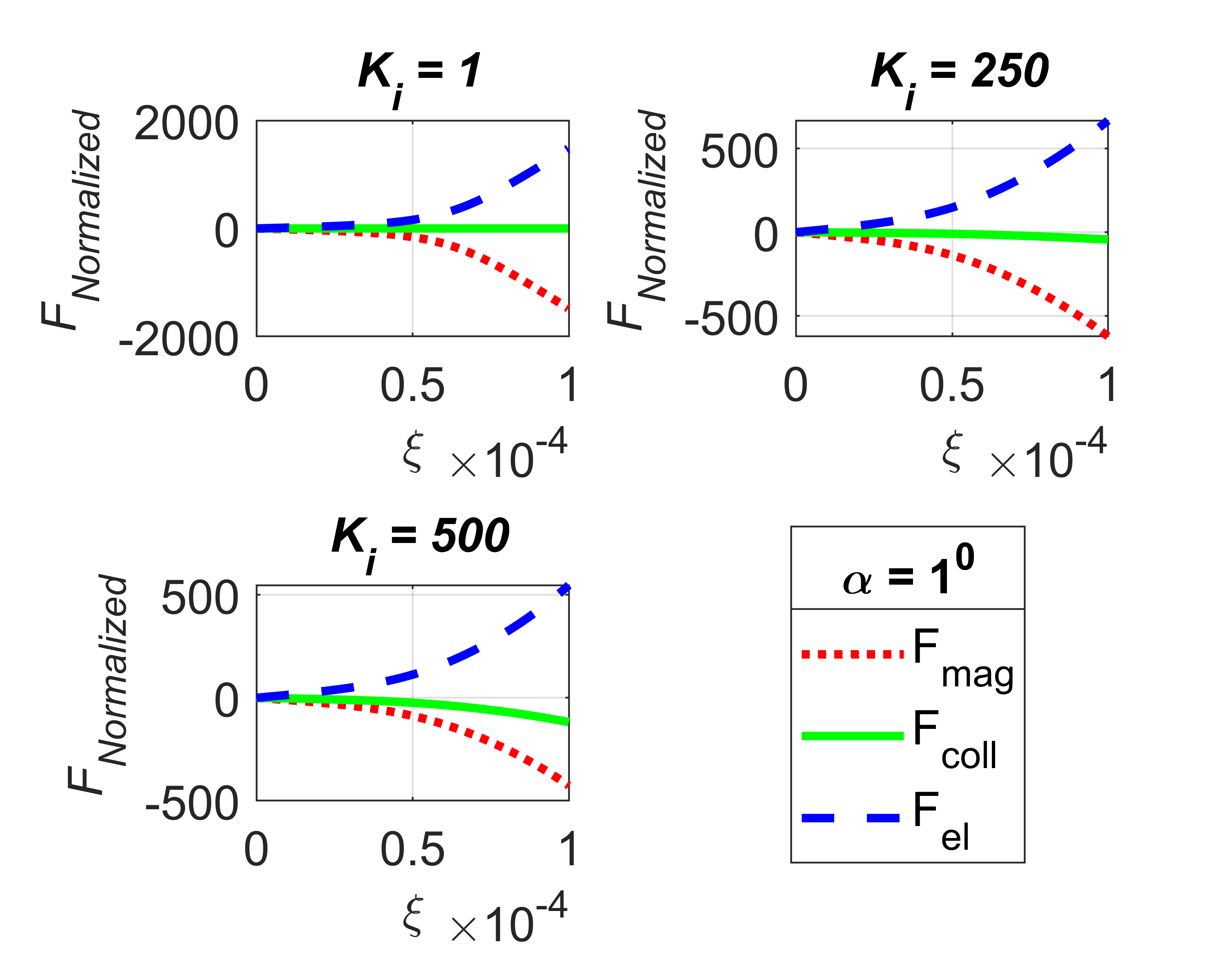}
    \end{minipage}
    \begin{minipage}[b]{0.48\textwidth}
    \includegraphics[width=1\textwidth]{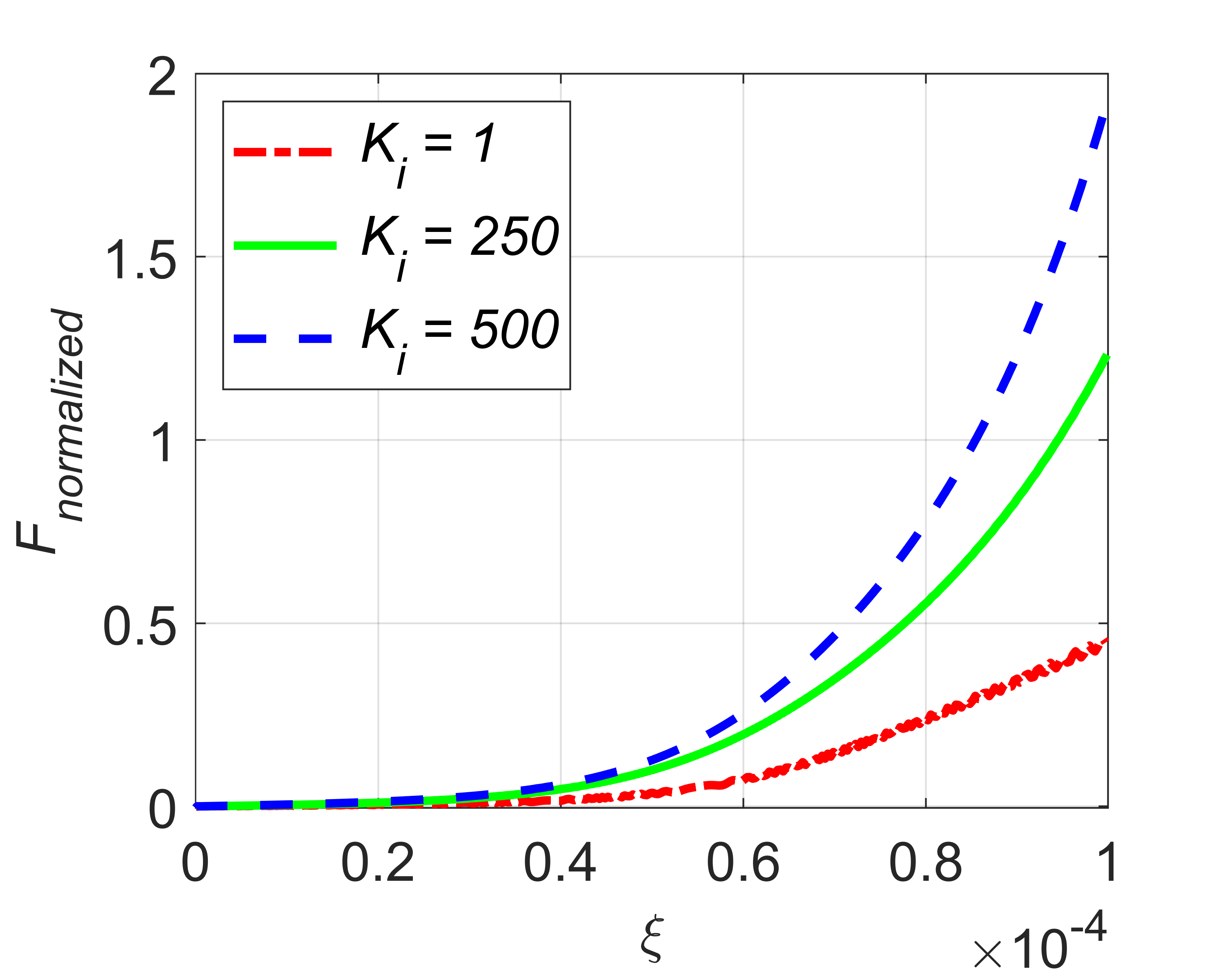}
    \end{minipage}
    \caption{Behaviour of various force fields (left) and the resultant force acting on the ions (right) for $\alpha = 1^0$ }
    \label{fig:20}
\end{figure}

\begin{figure}
    \centering
    \begin{minipage}[b]{0.48\textwidth}
    \includegraphics[width=1\textwidth]{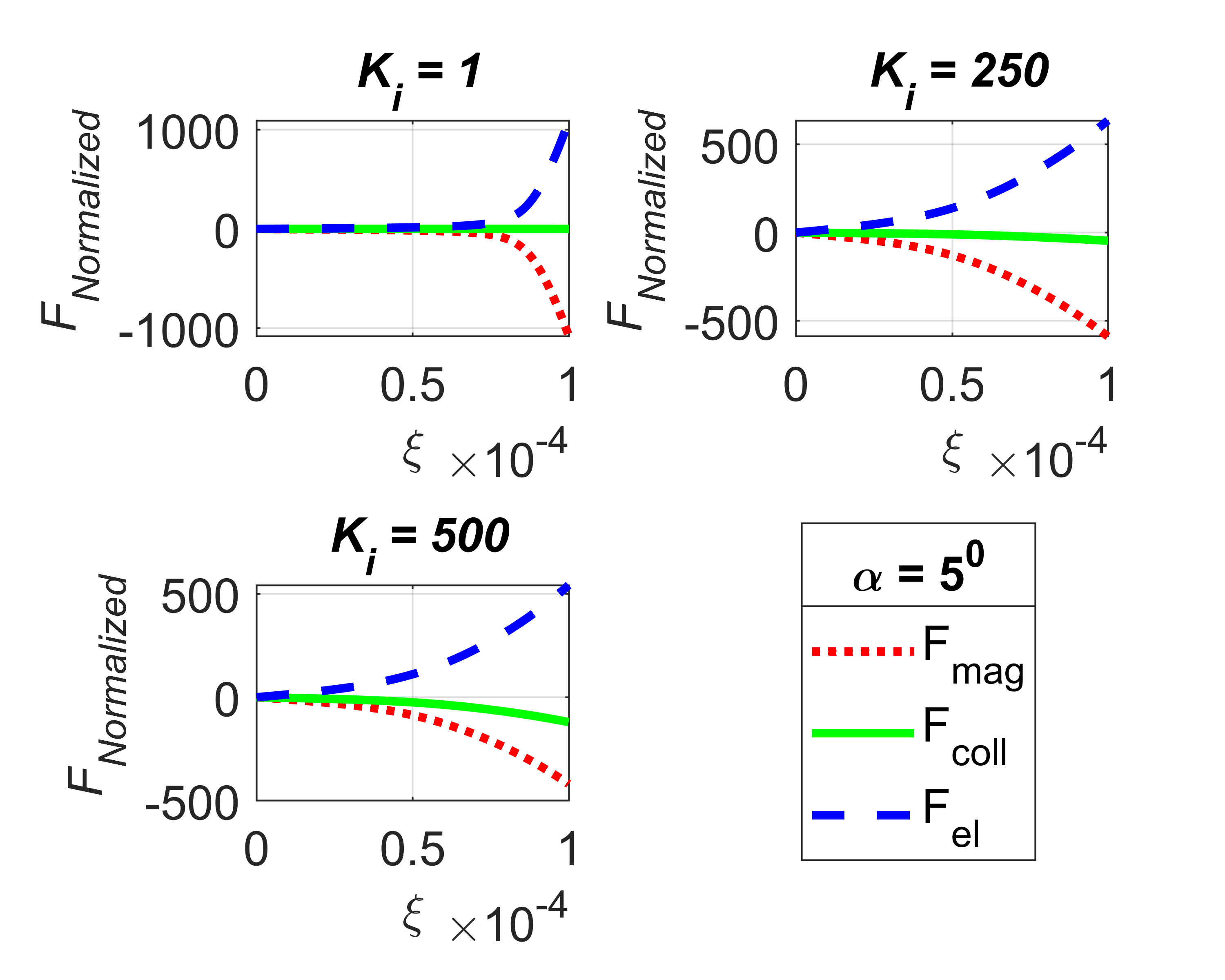}
    \end{minipage}
    \begin{minipage}[b]{0.48\textwidth}
    \includegraphics[width=1\textwidth]{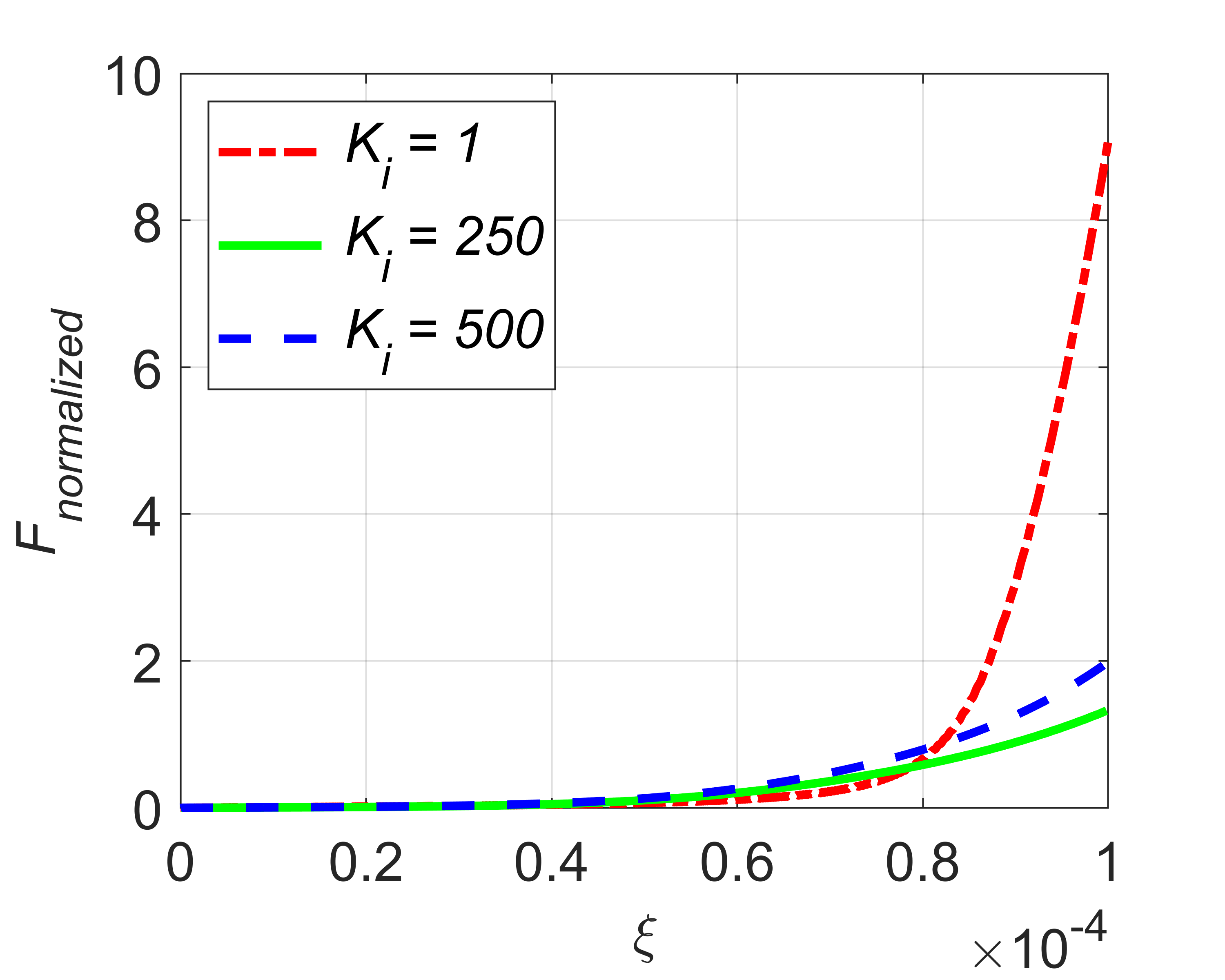}
    \end{minipage}
    \caption{Behaviour of various force fields (left) and the resultant force acting on the ions (right) for $\alpha = 5^0$ }
    \label{fig:21}
\end{figure}

Figure \ref{fig:16} and figure \ref{fig:17} displays the variations of ion flux and sheath thickness with $q$ for four different hot electron temperatures. It is observed that hot electron temperature has negligible effect on the ion flux. However, towards the low $q$ regime, the hot electron temperature affects the sheath thickness. As mentioned earlier, electrons with higher thermal energy can overcome the sheath potential barrier. Hence, for a fixed density of hot electrons, the total electron density inside the sheath will increase with the increase in electron thermal velocity, as observed in figure \ref{fig:17}.

\begin{figure}
    \centering
    \begin{minipage}[b]{0.48\textwidth}
    \includegraphics[width=1\textwidth]{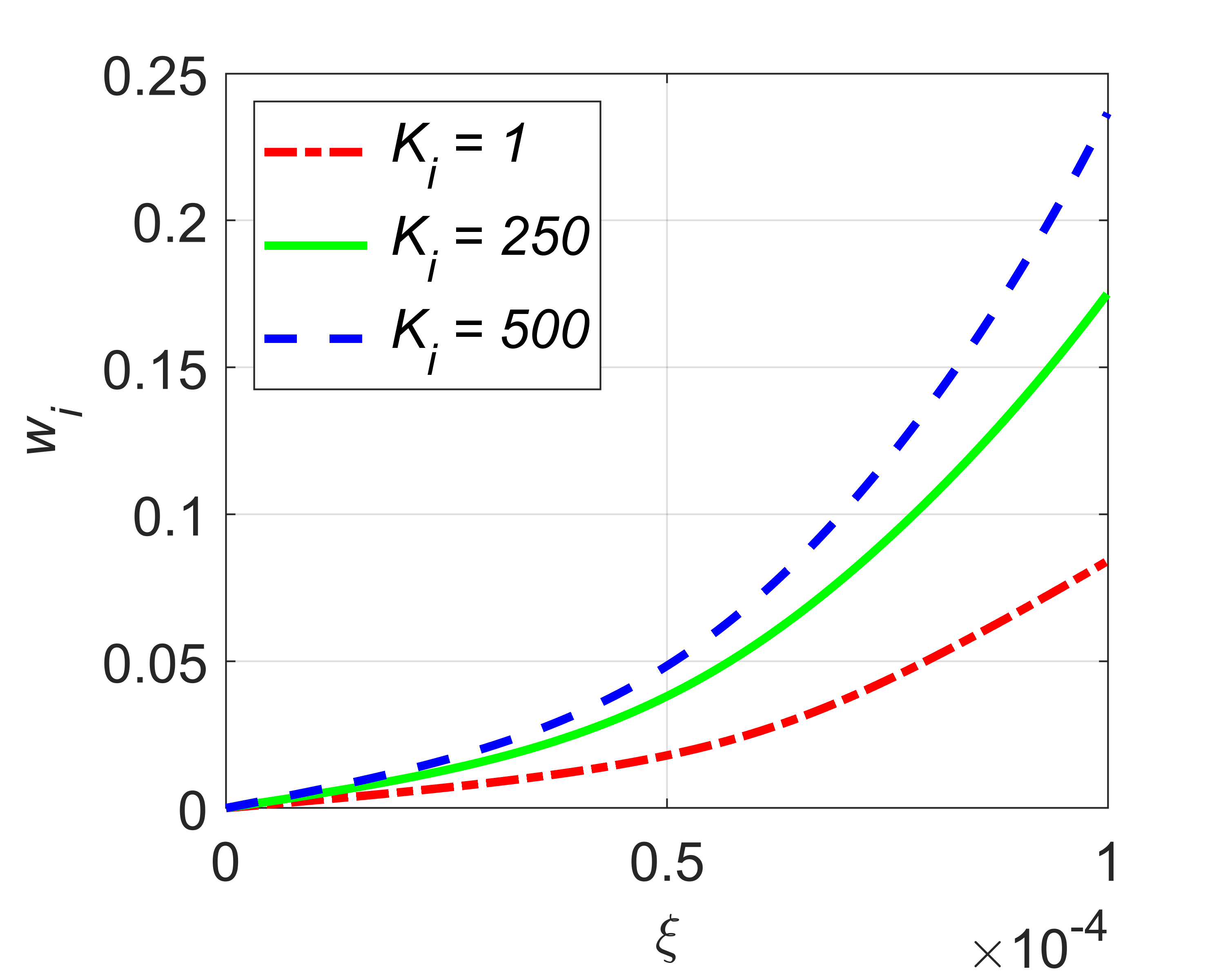}
    \end{minipage}
    \begin{minipage}[b]{0.48\textwidth}
    \includegraphics[width=1\textwidth]{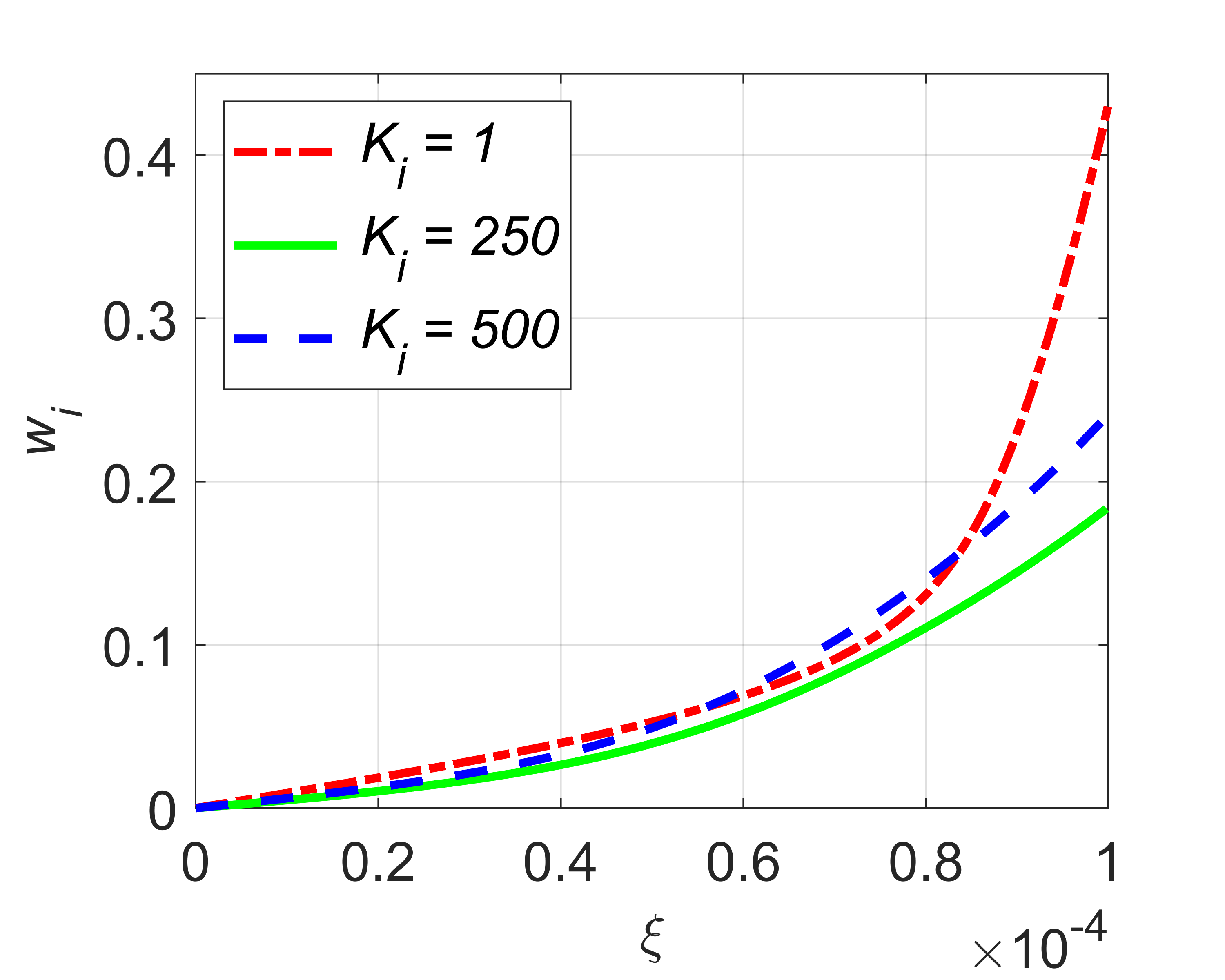}
    \end{minipage}
    \caption{Ion velocity (z-component) in the sheath for $\alpha=1^0$ (left) and $\alpha=5^0$ (right) in three collisional regimes}
    \label{fig:22}
\end{figure}

The response of ion flux and sheath width to the electron nonextensivity in different collisional regimes have been portrayed in figure \ref{fig:18} and figure \ref{fig:19} respectively. Substantial differences are observed in both ion flux and sheath thickness for three different values of collision parameters $K_i$. The rise in the ion flux with an increase in $K_i$ is an important finding of the study. Usually, collision tends to lower the ion velocity, and a consequential decrease in the ion flux is observed in the presence of an oblique magnetic field. However, the magnetic field is kept almost parallel to the wall in this particular study ($\alpha = 1^0$). This restricts the movement of the ions towards the plasma boundary. In this scenario, ion-neutral collision plays a noteworthy role by enabling cross-field diffusion of plasma particles towards the wall. An analysis of different forces controlling the ions movement is carried out for better understanding of the situation. Figure \ref{fig:20}
and figure \ref{fig:21} display the $z$-component of various forces and the resultant force acting on the ions  for different angle of inclination. In all the cases, the electrostatic and magnetic forces are opposite to each other. For $\alpha = 1^0$, the magnitude of all the forces decreases with the increase in $K_i$. However, the net force increases with $K_i$ for $\alpha = 1^0$ (figure \ref{fig:20}(right)), whereas it decreases with $K_i$ for $\alpha = 150$ (figure \ref{fig:21}(right)). As a consequence, the ion velocity increases with collision parameter for $\alpha = 1^0$ and decreases with collision for $\alpha = 5^0$ as shown in figure \ref{fig:22}. Therefore, a higher ion flux is observed for higher collision conditions when the magnetic field is nearly parallel to the wall.


\section{Conclusion}
A realistic range for the nonextensive parameter $q$ has been predicted for a low pressure magnetized plasma considering the sheath formation near a floating wall. At first, a comparative study has been carried out between a two-fluid model and a single fluid model with nonextensive electrons. The results of the comparative study reveal that for a magnetic field almost parallel to the wall ($\alpha=1^0$), the electron distribution considerably deviates from the Boltzmann distribution depending on the electron-neutral collisionality of the considered plasma. The range, $0.5\leq q \leq 1.0$, is found to be suitable. However, for $\alpha=5^0$, the electron distribution hardly deviates from the Boltzmann distribution. Hence, the angle of inclination of the magnetic field has a significant role in determining the electron distribution in plasmas. The results of the study have been used to investigate the effect of two-electron temperatures on the properties of a sheath in a magnetized plasma. The nonextensive distribution is considered to describe both the electron groups and found that ion flux to the wall and sheath width increases as the electron distribution deviates from the Boltzmann distribution. The fractional density of hot electrons has a profound effect on the sheath properties whereas the effect of their temperature is negligibly small. Moreover, the ion flux is found to increase with ion-neutral collision frequency for $\alpha=1^0$. This indicates that, for such a magnetic field configuration, cross-field diffusion might be a key factor in the formation of the sheath near the wall.


\section*{References}
\begin{thebibliography}{50}
    \bibitem{KU} Riemann K-U 1990  \textit{J. Phys. D: Appl. Phys.} \textbf{24} 493-518
    
    \bibitem{Bohm} Bohm D 1949 \textit{The Characteristics of Electrical Discharges in Magnetic
    Fields} (McGraw-Hill, New York)
    
    \bibitem{Val} Valentini H-B 1996 \textit{Phys. Plasmas} \textbf{3} 1459
    
    \bibitem{Chodu} Chodura R 1982 \textit{Phys. Fluids} \textit{25} 1628
    
    \bibitem{Hatami1} Hatami M M, Shokri B and Niknam A R 2008 \textit{J. Phys. D: Appl. Phys.} \textbf{42} 025204
    
    \bibitem{Stan} Stangeby P C 1995  \textit{Plasma Phys. Control. Fusion} \textbf{37} 1031
    
    \bibitem{Sheridan} Sheridan T E, Goeckner M J and Goree J 2003 \textit{J. Vac Sci. Technol. A} \textbf{9} 688
    
    \bibitem{Ike} Ikezawa S and Nakamura Y 1981 \textit{J. Phys. Soc. Jpn} \textbf{50} 962-967
    
    \bibitem{My_2} Sharma G, Deka K, Paul R, Adhikari S, Moulick R, Kausik S S and Saikia B K 2022 \textit{Plasma Sources Sci. Technol.} \textbf{31} 025013
    
    \bibitem{Shukla} Bharuthram R and Shukla P K 1986 \textit{Phys. Fluids} \textbf{29} 3214
    
    \bibitem{Gyr} Gyergyek T and Čerček M 2005 \textit{Contrib. Plasma Phys.} \textbf{45} 89-110
    
    
    \bibitem{Ou} Ou J, Xiang N, Gan C and Yang J 2013 \textit{Phys. Plasmas} \textbf{20} 063502
    
    \bibitem{My} Sharma G, Adhikari S, Moulick R, Kausik S S and Saikia B K 2020 \textit{Phys. Scr.} \textbf{95} 035605
    
    \bibitem{Vasy} Vasyliunas V M 1968 \textit{J. Geophys. Res.} \textbf{73} 2839
    
    \bibitem{Koen} Koen E J, Collier A B and Maharaj S K 2012 \textit{Phys. Plasmas} \textbf{19} 042102
    
    \bibitem{Jaw} Jaworski \textit{et al.} 2013 \textit{J. Nucl. Mater.} \textbf{438} S384-S387
    
    \bibitem{Kakati} Kakati B, Kausik S S, Saikia B K and Bandyopadhyay M 2011 \textit{Phys. Plasmas} \textit{18} 033705
    
    \bibitem{Kalita} Kalita D, Kakati B, Kausik S S, Saikia B K and Bandyopadhyay M, 2019 \textit{J. Plasma Phys} \textbf{85} 905850402
    
    \bibitem{Chen} Chen L, Jin D, Tan X, Dai J, Cheng L and Hu S, 2010 \textit{Vacuum} \textbf{85} 622
    
    \bibitem{Liu} Liu J M, De Groot J S, Matte J P, Johnston T W and Drake R P 1994 \textit{Phys. Rev. Lett.} \textbf{72} 2717
     
    \bibitem{Tsa} Tsallis C 1988 \textit{J. Stat. Phys.} \textbf{52} 479
    
    \bibitem{Fran} Franklin R N 2011 \textit{J. Plasma Physics} \textbf{78} 21-24
    
    \bibitem{Tsha} Tskhakaya D 2017 \textit{Plasma Phys. Control. Fusion} \textbf{59} 114001
    
    \bibitem{Hatami2} Hatami M M 2015 \textit{Phys. Plasmas} \textbf{22} 013508
    
    \bibitem{Safa} Safa N N, Ghomi H and Niknam A R 2015 \textit{J. Plasma Physics} \textbf{81} 905810303
    
    \bibitem{QH} Qiu H, Song H, and Liua S 2014 \textit{Phys. Plasmas} \textbf{21} 062310
    
    \bibitem{Bas} Basnet S and Khanal R 2019 \textit{Phys. Plasmas} \textbf{26} 043516
    
    \bibitem{Dima} Borgohain D R, Saharia K and Goswami K S 2016 \textit{Phys. Plasmas} \textbf{23} 122113
    
    \bibitem{Dima2} Borgohain D R and Saharia K 2019 \textit{Indian J. Phys } \textbf{93}(1) 107–114
    
    \bibitem{Safa2} Safa N N, Ghomi H and Niknam A R 2014 \textit{Phys. Plasmas} \textbf{21} 082111
    
    \bibitem{Mou} Moulick R, Garg A and Kumar M 2021 \textit{Contrib. Plasma Phys.} \textbf{61} e202100047
    
    \bibitem{Bog} Boghosian B M 1996 \textit{Phys. Rev. E} \textbf{53} 4754 
    
    \bibitem{Du} Du J 2006 \textit{Europhys. Lett.} \textbf{75} 861
    
    \bibitem{Qiu} Qiu \textit{et al.} 2020 \textit{Phys. Rev. E} \textbf{101} 043206
    
    \bibitem{kishor} Deka K, Adhikari S, Moulick R, Kausik S S and Saikia B K 2021 \textit{Phys. Scr.} \textbf{96} 075606
    
    \bibitem{maso} Masoudi S F, Esmaeili S S and Jazavandi S 2009 \textit{Vacuum} \textbf{382} 386
    
    \bibitem{RM} Moulick R, Adhikari S and Goswami K S 2019 \textit{Phys. Plasmas} \textbf{26} 043512
    
    \bibitem{RN} Franklin R N 2012 \textit{J. Plasma Phys.} \textbf{78} 21–24
    
    \bibitem{Hat} Hatami M M 2021 \textit{Sci. Rep.} \textbf{11} 9531
    
    \bibitem{PC} Stangeby P C 2012 \textit{Nucl. Fusion} \textbf{52} 083012
    
    \bibitem{Tsa2} Tskhakaya D D, Shukla P K, Eliasson B and Kuhn S 2005 \textit{Phys. Plasmas} \textbf{12}  103503 
    
    
    
    
    
    
    
   
    
\end{thebibliography}
\end{document}